\documentclass[nonacm,sigconf]{acmart}

\AtBeginDocument{%
  \providecommand\BibTeX{{%
    \normalfont B\kern-0.5em{\scshape i\kern-0.25em b}\kern-0.8em\TeX}}}

\setcopyright{acmcopyright}
\copyrightyear{2023}
\acmYear{2023}
\acmDOI{XXXXXXX.XXXXXXX}

\acmConference[SC'23]{International Conference for High Performance Computing, Networking, Storage and Analysis}{Nov 12--17,
  2023}{Denver, CO}
\acmBooktitle{SC'23: International Conference for High Performance Computing, Networking, Storage and Analysis,
 Nov 12--17, 2023, Denver, CO} 
\acmPrice{15.00}
\acmISBN{978-1-4503-XXXX-X/18/06}

\usepackage{varwidth}
\usepackage{subcaption}
\usepackage{listings}
\usepackage{xcolor}
\usepackage{enumitem}

\definecolor{codegreen}{rgb}{0,0.6,0}
\definecolor{codegray}{rgb}{0.5,0.5,0.5}
\definecolor{codepurple}{rgb}{0.58,0,0.82}
\definecolor{backcolour}{rgb}{0.95,0.95,0.92}

\lstdefinestyle{mystyle}{
    commentstyle=\color{codegreen},
    keywordstyle=\color{magenta},
    numberstyle=\tiny\color{codegray},
    stringstyle=\color{codepurple},
    basicstyle=\ttfamily\footnotesize,
    breakatwhitespace=false,         
    breaklines=true,                 
    captionpos=b,                    
    keepspaces=true,                 
    numbers=left,                    
    numbersep=5pt,                  
    showspaces=false,                
    showstringspaces=false,
    showtabs=false,                  
    tabsize=2
}
\DeclareCaptionLabelFormat{andtable}{#1~#2  \&  \tablename~\thetable}

\usepackage{tikz}
\newcommand\encircle[1]{%
  \tikz[baseline=(X.base)]
    \node (X) [font=\fontsize{10}{0}\selectfont,draw, shape=circle, inner sep=0, fill=black, text=white] {\strut #1};%
}

\usepackage{tikz}
\newcommand\encircles[1]{%
  \tikz[baseline=(X.base)]
    \node (X) [font=\fontsize{8}{0}\selectfont,draw, shape=circle, inner sep=0, fill=black, text=white] {\strut #1};%
}

\def\BibTeX{{\rm B\kern-.05em{\sc i\kern-.025em b}\kern-.08em
    T\kern-.1667em\lower.7ex\hbox{E}\kern-.125emX}}
    
\begin{document}

\title{VENOM: A Vectorized N:M Format for Unleashing the Power of Sparse Tensor Cores}

\author{Roberto L.~Castro}
\email{roberto.lopez.castro@udc.es}
\orcid{0000-0001-5493-0287}
\authornote{Corresponding author: Roberto L.~Castro (roberto.lopez.castro@udc.es), Universidade da Coru\~{n}a, CITIC, Computer Architecture Group, 15071 A Coru\~{n}a, Spain}
\affiliation{%
  \department{CITIC}
  \institution{Universidade da Coru\~{n}a}
  \streetaddress{Campus de Elvi\~{n}a}
  \city{A Coru\~{n}a}
  \country{Spain}
  \postcode{15071}
}

\author{Andrei Ivanov}
\email{anivanov@inf.ethz.ch}
\orcid{0009-0007-9487-9990}
\affiliation{%
  \department{Department of Computer Science}
  \institution{ETH Z\"{u}rich}
  \city{Z\"{u}rich}
  \country{Switzerland}
}

\author{Diego Andrade}
\email{diego.andrade@udc.es}
\orcid{0000-0001-5670-7425}
\affiliation{%
  \department{CITIC}
  \institution{Universidade da Coru\~{n}a}
  \streetaddress{Campus de Elvi\~{n}a}
  \city{A Coru\~{n}a}
  \country{Spain}
  \postcode{15071}
}

\author{Tal Ben-Nun}
\email{talbn@inf.ethz.ch}
\orcid{0000-0002-3657-6568}
\affiliation{%
  \department{Department of Computer Science}
  \institution{ETH Z\"{u}rich}
  \city{Z\"{u}rich}
  \country{Switzerland}
}

\author{Basilio B.~Fraguela}
\email{basilio.fraguela@udc.es}
\orcid{0000-0002-3438-5960}
\affiliation{%
  \department{CITIC}
  \institution{Universidade da Coru\~{n}a}
  \streetaddress{Campus de Elvi\~{n}a}
  \city{A Coru\~{n}a}
  \country{Spain}
  \postcode{15071}
}

\author{Torsten Hoefler}
\email{htor@inf.ethz.ch}
\orcid{0000-0002-1333-9797}
\affiliation{%
  \department{Department of Computer Science}
  \institution{ETH Z\"{u}rich}
  \city{Z\"{u}rich}
  \country{Switzerland}
}

\renewcommand{\shortauthors}{Castro R.L., et al.}

\begin{abstract}
The increasing success and scaling of Deep Learning models demands higher computational efficiency and power. Sparsification can lead to both smaller models as well as higher compute efficiency, and accelerated hardware is becoming available.
However, exploiting it efficiently requires kernel implementations, pruning algorithms, and storage formats, to utilize hardware support of specialized sparse vector units. 
An example of those are the NVIDIA’s Sparse Tensor Cores (SPTCs), which promise a $2\times$ speedup. However, SPTCs only support the 2:4 format, limiting achievable sparsity ratios to $50\%$. We present the V:N:M format, which enables the execution of arbitrary N:M ratios on SPTCs. To efficiently exploit the resulting format, we propose \emph{Spatha}, a high-performance sparse-library for DL routines. We show that Spatha achieves up to $37\times$ speedup over cuBLAS. We also demonstrate a second-order pruning technique that enables sparsification to high sparsity ratios with V:N:M and little to no loss in accuracy in modern transformers.
\end{abstract}






\maketitle

\section{Introduction}
The rapid progress of Deep Learning (DL) is revolutionizing Artificial Intelligence (AI) in areas such as Natural Language Processing (NLP). Large Language Models (LLMs) are at the forefront of modern NLP systems~\cite{DBLP:conf/naacl/DevlinCLT19,NIPS2017_3f5ee243}; however, their massive growth has led to unprecedented computational requirements~\cite{hestness2017deep,kaplan2020scaling,NEURIPS2020_1457c0d6,openai}. As a result, training transformers has become a dominant task in DL, with costs reaching millions of dollars and significant energy and carbon emissions~\cite{strubell-etal-2019-energy}. Therefore, it is critical to improve their inference and training performance. One of the most widely used techniques for this purpose is network pruning~\cite{10.5555/3546258.3546499}, which removes the less significant weights to produce simpler and compressed, yet accurate models.

There is a plethora of pruning algorithms and sparse formats focused on accelerating tensor operations such as matrix-matrix multiplications (MMMs) by means of specialized hardware like Tensor Core Units (TCUs)~\cite{9139835}. 
While these algorithms and formats reduce the number of arithmetic operations and memory usage compared to their dense counterparts, achieving significant speedup on these accelerators while maintaining model accuracy is challenging~\cite{data-movement-is-all-you-need}. 
Semi-structured pruning can yield practical speedups at moderate sparsity levels (e.g., $80-90\%$)~\cite{10.1145/3559009.3569691,10.1145/3458817.3476182,10.5555/3571885.3571934}. However, the irregularity of the sparse input matrices still limits performance and makes difficult to reach the theoretical peak considering the reduction of the number of arithmetic operations~\cite{gale2020sparse}.

Last generations of NVIDIA GPUs include Sparse Tensor Cores (SPTCs) that are specifically designed for sparse computation~\cite{mishra2021accelerating}. SPTCs promise to accelerate math operations by up to $2\times$ at $50\%$ sparsity. The data layout proposed to use SPTCs imposes strict constraints (i.e., 2:4 format, where every consecutive 4 elements have 2 nonzero values), but it reduces the irregularity of the sparse input w.r.t. other performance-aware sparse formats (e.g., vector-wise, block-wise). This makes the N:M format very suitable to execute on GPUs since it favors key aspects of the execution of tensor operations such as inter- and intra-warp load balance. However, there is an important limitation related to the usage of SPTCs and the 2:4 format: recent models like LLMs commonly have hundreds of millions to trillions of parameters, making it feasible to prune them to higher sparsity ratios with little or no loss in accuracy~\cite{oBERT}. Unfortunately, there is currently no hardware support for executing arbitrary N:M formats with higher compression ratios, which limits the total achievable speedup. 

Recent research has explored the N:M format~\cite{10.1145/3572848.3577500,frantar2023sparsegpt}. However, these investigations have been limited to a theoretical perspective, such as network pruning, or have relied on CPU implementations due to a lack of hardware support for alternative N:M patterns on GPUs. To address these limitations, we propose the Vectorized N:M format, which we refer to as V:N:M\footnote{Pronounced ``venom''}. This format introduces an abstraction layer over SPTCs, enabling the execution of alternative N:M formats and arbitrary sparsity ratios. The vectorization aspect is derived from the selection of vertical vectors of elements that are stacked together to provide the row-wise N:M pattern. This approach enables the conversion from generic N:M formats to the 2:4 that is accepted by SPTCs. To efficiently exploit the benefits of the V:N:M format, we propose Spatha\footnote{SParse linear Algebra rouTines for High-performance Applications. The name is motivated by the analogy with the Cutlass library, with the accent on sparse computation - a sharp and efficient tool to cut through the complexity of sparse routines}, a template-based library dedicated to general matrix-matrix multiplication on half precision where one of the operands is sparse (SpMM). Spatha serves as an open-source alternative to cuSparseLt~\cite{cusparselt} and removes its 2:4 restriction. The main contributions of this paper are:

\begin{itemize}[left=3.5pt]
    \item A new sparse matrix format V:N:M which enables arbitrary N:M patterns on SPTCs.
    \item Highly optimized SpMM kernels to efficiently exploit the V:N:M format. Specifically, we propose a template-based implementation that can be tuned depending on the input dynamics, such as GEMM size or the V:N:M format configuration.
    \item A second-order pruning technique tailored for the V:N:M format and scalable to the dimensionality of LLMs. This technique allows the sparsification to high sparsity ratios with little to no loss in accuracy (e.g., $\sim2\%$ drop in BERT F1 score on the SQuAD dataset with 2:16 sparsity), which is required for the full exploitation of the V:N:M format.
    \item Spatha achieves unprecedented speedups w.r.t. its dense counterpart versions (e.g., cuBLAS) yielding up to $37\times$ faster MMMs on matrices extracted from real-world DL models. Furthermore, Spatha implementation provides speedups of up to $1.38\times$ over the vendor library for 2:4 sparsity, cuSparseLt.
    \item For end-to-end sparse LLMs inference, Spatha shows a GEMM time reduction of $11\times$ at 2:32 sparsity on real-world models such as GPT-3.
\end{itemize}

The source code of VENOM is available at \url{https://github.com/UDC-GAC/venom}.

\section{BACKGROUND}
This section presents the technical background of the paper,
covering network pruning techniques and the Sparse Tensor Cores of NVIDIA GPUs.
\subsection{Network pruning}
In DL, pruning is a technique used to reduce memory usage, which can also reduce the computational load when combined with compressed storage formats and efficient sparse kernels. Pruning techniques can be categorized based on various criteria, such as the pruning strategy employed, or the granularity of the pruning.

 Pruning schemes are often based on weight saliency metrics, which directly correlate with the expected impact on accuracy when those weights are removed from the network.
 Various methods exist to select the candidate weights for removal, including magnitude pruning~\cite{kurtic2022gmp}, which selects weights with lower absolute values, and gradient-based methods that use the gradient applied to each weight to identify those that are trending towards to zero faster. Within the gradient-based methods, we can find first-order techniques based on the first-derivative information~\cite{sanh2020movement,pmlr-v162-zhang22ao}, and second-order ones~\cite{oBERT,LeCun1989OptimalBD,frantar2021mfac}, which pursue to find the set of weights whose removal will generate a minimum loss increase in the network. Second-order methods have proven to be effective in pruning convolutional networks in the past, but they have recently been optimized for Large Language Models (LLMs)~\cite{oBERT}.

As for the granularity of the pruning, unstructured methods~\cite{han2016deep} remove weights individually, with gradual magnitude pruning (GMP) being the most commonly used variant~\cite{gale2019state}. On the other end of the granularity spectrum, structured methods~\cite{NEURIPS2019_2c601ad9,voita2019analyzing} prune complete components like layers, or heads, in the case of transformers networks\cite{wang2019structured}. In between, semi-structured methods prune groups of weights. These latter methods aim to balance performance and accuracy by defining specific formats that promote the exploitation of the underlying hardware more efficiently. These methods often imply the usage of tailored compressed storage formats and custom kernels~\cite{lagunas2021block,gale2020sparse}. 
The N:M format, which enables the use of Sparse Tensor Cores (SPTCs) in NVIDIA GPUs, can be classified in this last group.

\subsection{Sparse Tensor Cores of NVIDIA GPUs}

The CUDA programming model organizes GPU kernels into three granularity levels: thread-blocks, warps, and threads. A thread block is composed of a set of warps, with warps being the basic scheduling unit in CUDA. Each warp consists of $32$ threads.

NVIDIA GPUs consist of an array of Streaming Multiprocessors (SMs), with all SMs sharing the L2 cache, and a DRAM memory, also called Global Memory (GMEM). Each SM is divided in processing blocks, each one having a Register File (RF), a warp scheduler, and an L0 instruction cache. All the processing blocks within an SM share a L1 cache, which is 
partially used as Shared Memory (SMEM). Each processing block is also equipped with four types of units: Floating-Point Units (FPU), Tensor-Core Units (TCU), Int Units (ALU) and Special Function Units (SFU).

\begin{figure}[ht]
    \centerline{\includegraphics[width=\linewidth]{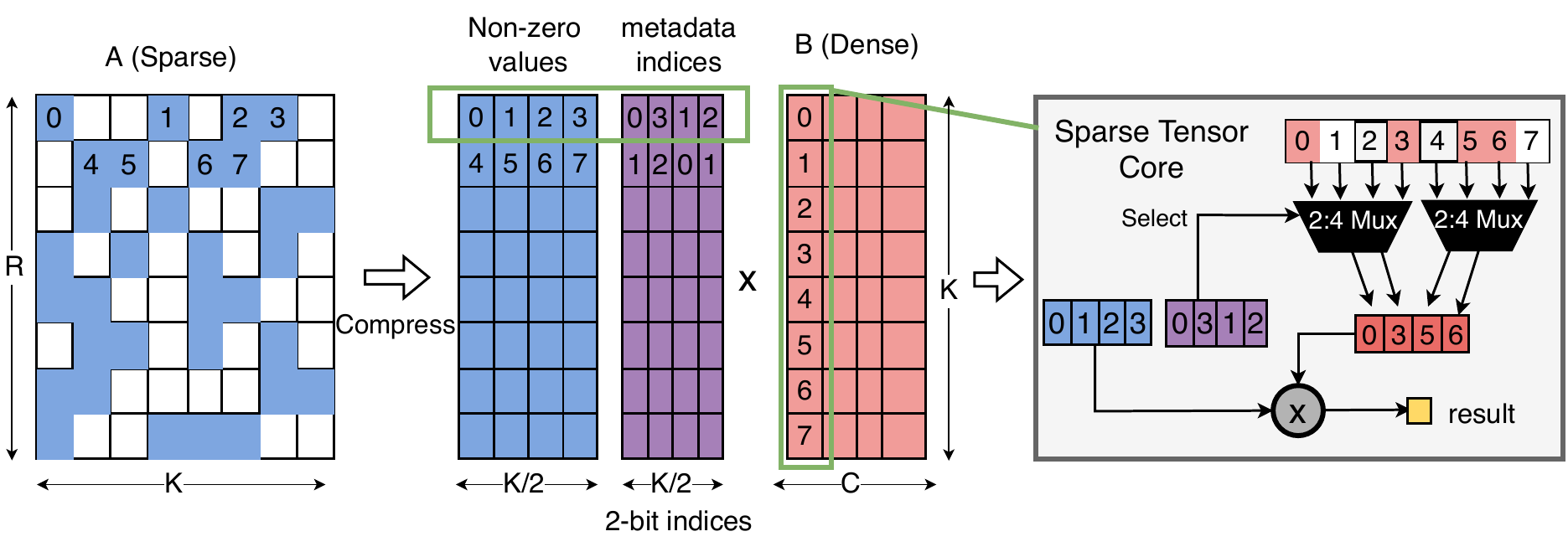}}
    \caption{The 2:4 format and its mapping to SPTCs}
    \label{fig:nvidia_layout}
\end{figure}

Last generations of NVIDIA GPUs have extended their TCUs to also handle row-wise 2:4 sparsity. These updated TCUs include hardware support for sparse computation, and are referred to as Sparse Tensor Cores (SPTCs). To exploit SPTCs, the first argument in tensor operations must be stored in NVIDIA's N:M sparse format, where $N$ represents the maximum number of non-zero elements in a block of $M$ values.
Figure~\ref{fig:nvidia_layout} illustrates this format.
The left side of the figure shows an uncompressed sparse matrix following the row-wise 2:4 pattern. The compression of that $R \times K$ matrix requires two structures: (1) a $R\times K/2$ matrix representing the values of the non-zero elements, and (2) a metadata structure which contains the position of each nonzero value within each group of $4$ values. Finally, Figure~\ref{fig:nvidia_layout}, right side, illustrates the mapping of a 2:4 sparse operation onto SPTCs. Notice that the metadata structure is also used by the hardware to select the corresponding elements in the dense matrix $B$ and perform the Matrix Multiply-Accumulate (MMA) operation.

\begin{table}[ht]
\begin{center}
\begin{tabular}{ ccc } 
 \toprule
 Precision & Format & Supported shapes     \\ \midrule
 fp32           & 1:2 & $k8, k16$   \\ 
 \textbf{\emph{half (fp16)}} & \textbf{\emph{2:4}} & k32, k16  \\ 
 uint8          & 2:4 & $k32, k64$  \\
 uint4          & 2:4 & $k64, k128$ \\
 \bottomrule
\end{tabular}
\caption{Matrix Shapes for \emph{mma.sp} on SPTCs. $M$ and $N$ dimensions are fixed to $16$ and $8$, respectively ($m16n8$)}
\label{tab:mma_shapes}
\end{center}
\end{table}

SPTCs can be accessed in CUDA using the NVPTX API which includes the \emph{mma.sp} instruction. 
SPTCs support various shapes of this instruction depending on the data precision (Table~\ref{tab:mma_shapes}).
This instruction multiplies a $m \times k$ matrix by a $k \times n$ matrix, where $m=16$, $n=8$ are fixed dimensions, and $k$ represents the sparsified dimension which can vary in size.
This paper focuses on half precision kernels. 
Instruction shapes define the sizes of the left-hand-side (LHS) and the right-hand-side (RHS) operands as inputs to TCUs. For example, $k=32$ implies that the LHS operand has a shape of $m \times k=16 \times 32$
while the RHS is $k\times n=32 \times 8$. It is important to note that the LHS is $50\%$ sparse, meaning that its real size will be $16 \times \boldsymbol{16} (32/2)$. NVIDIA's notation for this instruction is $m16n8k32$.

\section{The V:N:M format}

This section presents the new V:N:M format, which enables pruning to arbitrary N:M ratios retaining the use of SPTCs, which are designed to support only 2:4 patterns natively.

\begin{figure}[ht]
    \centerline{\includegraphics[width=\linewidth]{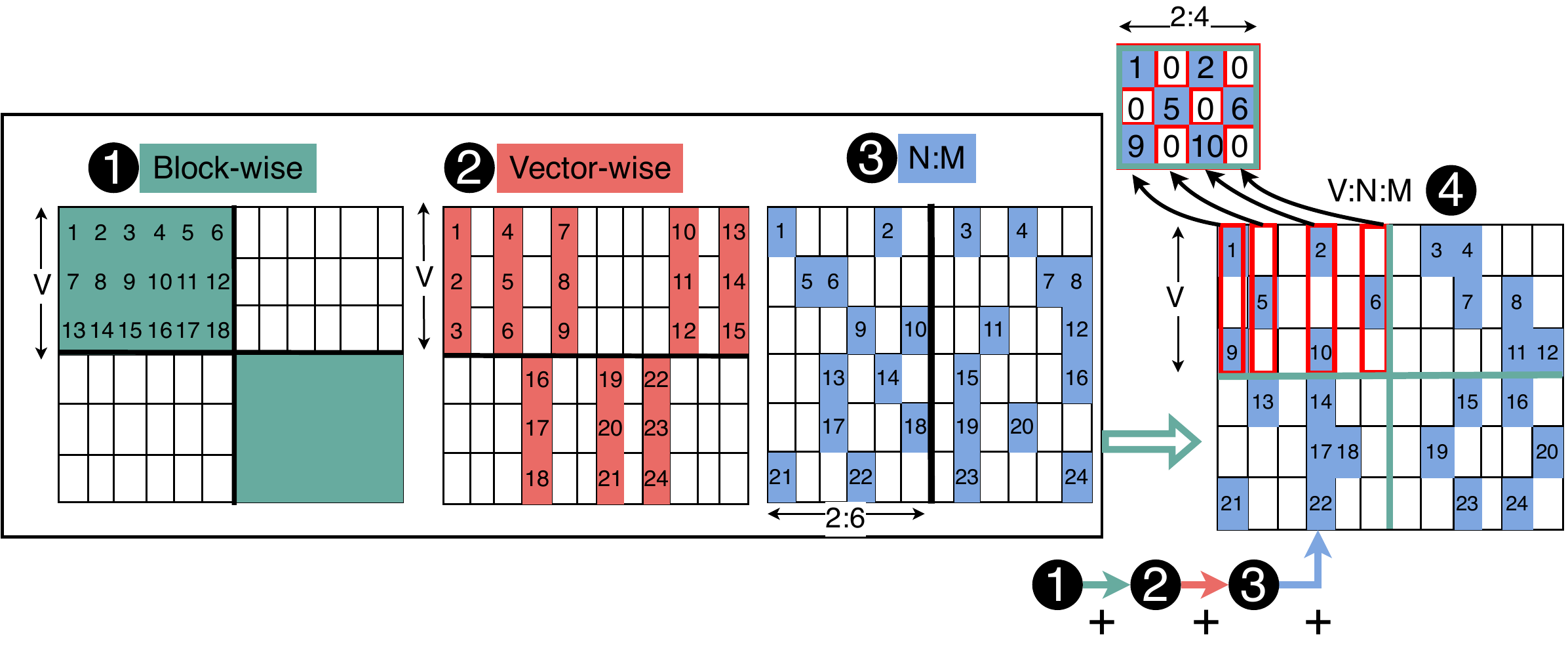}}
    \caption{The V:N:M pruning procedure}
    \label{fig:overview}
\end{figure}

Sparse compression formats are of great significance in many HPC areas other than DL. However, the characteristics of the sparse matrices in DL workloads differ from those in other areas in several aspects~\cite{gale2020sparse}: (1) the sparsity level is generally much lower, (2) the number of non-zeros per row is higher and (3) the load imbalance is more pronounced. 
To address these challenges, ad-hoc solutions for DL workloads have been developed in two different planes: compression formats and pruning techniques, often interlinked. They seek the efficient exploitation of the hardware during the execution of tensor operations in DL workloads. 

A new area of research is focused on enhancing control over the distribution of non-zero elements in sparse matrices. This involves, for example, selecting 2D dense groups with size $v \times v$ (Figure~\ref{fig:overview},~\encircle{1}) or 1D groups of length $v$, either row-wise or column-wise~\encircle{2}. The aim is to create sparse matrices that are more regular, making them more suitable for efficient execution on GPUs. Block-based pruning techniques (\encircle{1} and~\encircle{2}) are particularly useful on improving data reuse on L1 cache or registers during the multiplication of sparse matrices. Furthermore, optimized sparse formats, which compress their data, can be designed to facilitate traversal for the access patterns that arise during matrix multiplication~\cite{10.1145/331532.331562,10.5555/3571885.3571934}.

On the one hand,~\encircle{1} can be overly aggressive in dropping blocks of elements, leading to a significant reduction in accuracy as the sparsity level increases. On the other hand,~\encircle{2} offers more flexibility and enables higher sparsification ratios. However, using small vector lengths is a limiting factor to prevent accuracy loss (e.g., $v\leq 8$). Furthermore, in these approaches, the different number of elements per row can generate load imbalance and inherent negative effects such as thread divergence, inefficient memory transactions and low occupancy ratios. 

The N:M format~\encircle{3} provides an alternative that overcomes most of the weaknesses of other performance-aware methods. Moreover, NVIDIA GPUs recently included hardware support for this format, but it is limited to 2:4.
This paper introduces the new V:N:M format~\encircle{4} which combines block-wise storage, and vector-wise and N:M pruning to enable the exploitation of SPTCs for arbitrary N:M patterns, leveraging higher compression ratios and reducing further the number of arithmetic operations required in MMMs.

\begin{figure}[ht]
    \centerline{\includegraphics[width=0.6\linewidth]{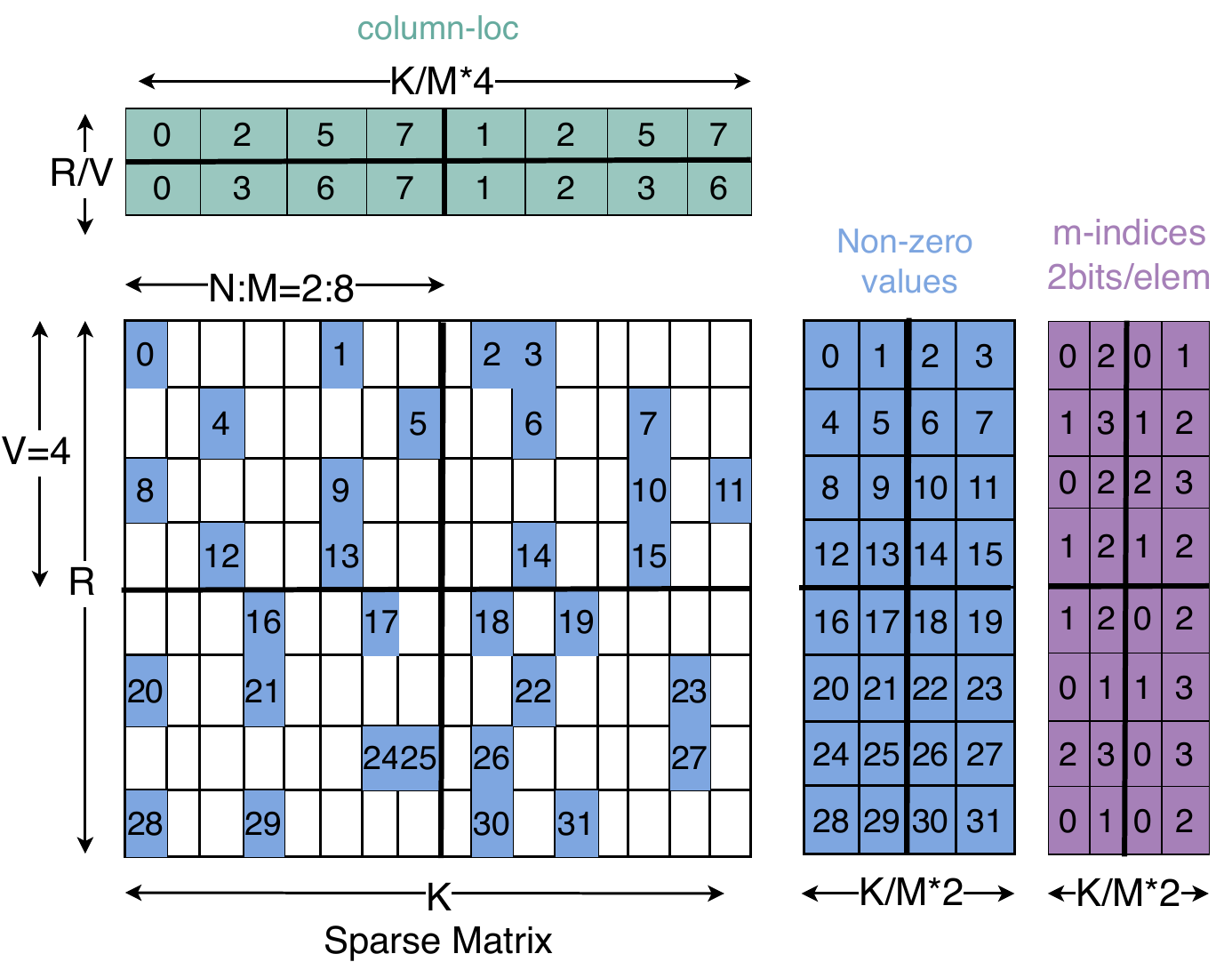}}
    \caption{The V:N:M compression format}
    \label{fig:n_m_v}
\end{figure}

Figure~\ref{fig:overview}, illustrates how this approach starts by partitioning the original dense matrix in blocks of $V\times M$ elements (block-wise).
Then, the four most significant columns of each block are selected (vector-wise pruning), and for each row of four elements in a block, the two most meaningful weights are kept (2:4 pruning).
These two levels of pruning (vector-wise and N:M) enable the exploitation of SPTCs for matrices with arbitrary levels of sparsity, as the vector-wise pruning stage diversifies the sparsity level, and N:M pruning imposes the restrictions required later by SPTCs. 
That is, in~\encircle{4}, the SPTC vector is 2:4, but it belongs to a 6-columns row, where $2$ columns were fully pruned. It is actually an implementation of a 2:6 sparsity pattern that it is mapped onto SPTCs as the required 2:4.

Finally, the data is represented using a new block-wise compression format shown in Figure~\ref{fig:n_m_v}.
As for the NVIDIA 2:4 layout (Figure~\ref{fig:nvidia_layout}), the format requires an array with the non-zero values, and a 2-bit metadata index per non-zero (\emph{m-indices}).
Notice that now, each 2-bit metadata index refers to one of the $4$ columns that we have selected in each block and not to each column of the original dense input matrix (see~\encircle{4} in Figure~\ref{fig:overview}). Furthermore, the size of these two structures depends on the M value, more specifically their shape now is $R\times K/M\times2$.
This format requires a third structure~\emph{column-loc} of size $R/V \times K/M\times4$, that indicates which $4$ columns (out of $M$) of each block were selected in the vector-wise pruning stage.

\section{Spatha: A High-Performance Sparse Library for Sparse MMM}

This section provides an in-depth description of the sparse kernel implementation associated to the V:N:M format, Spatha. The Sparse Matrix-Matrix multiplication (SpMM) is an important workload in DL that serves as the sparse counterpart to Matrix-Matrix Multiplication (MMM). This routine is widely used in various components of modern DL models. For instance, in the forward pass of a pruned model, the sparse weight matrix is multiplied by a dense activation matrix. Similarly, in transformers, the self-attention operation is performed by multiplying a sparse attention weight matrix by a dense one. Thus, optimizing this routine is crucial to improve the efficiency and the performance of our models.

\begin{figure}[ht]
    \centerline{\includegraphics[width=0.9\linewidth]{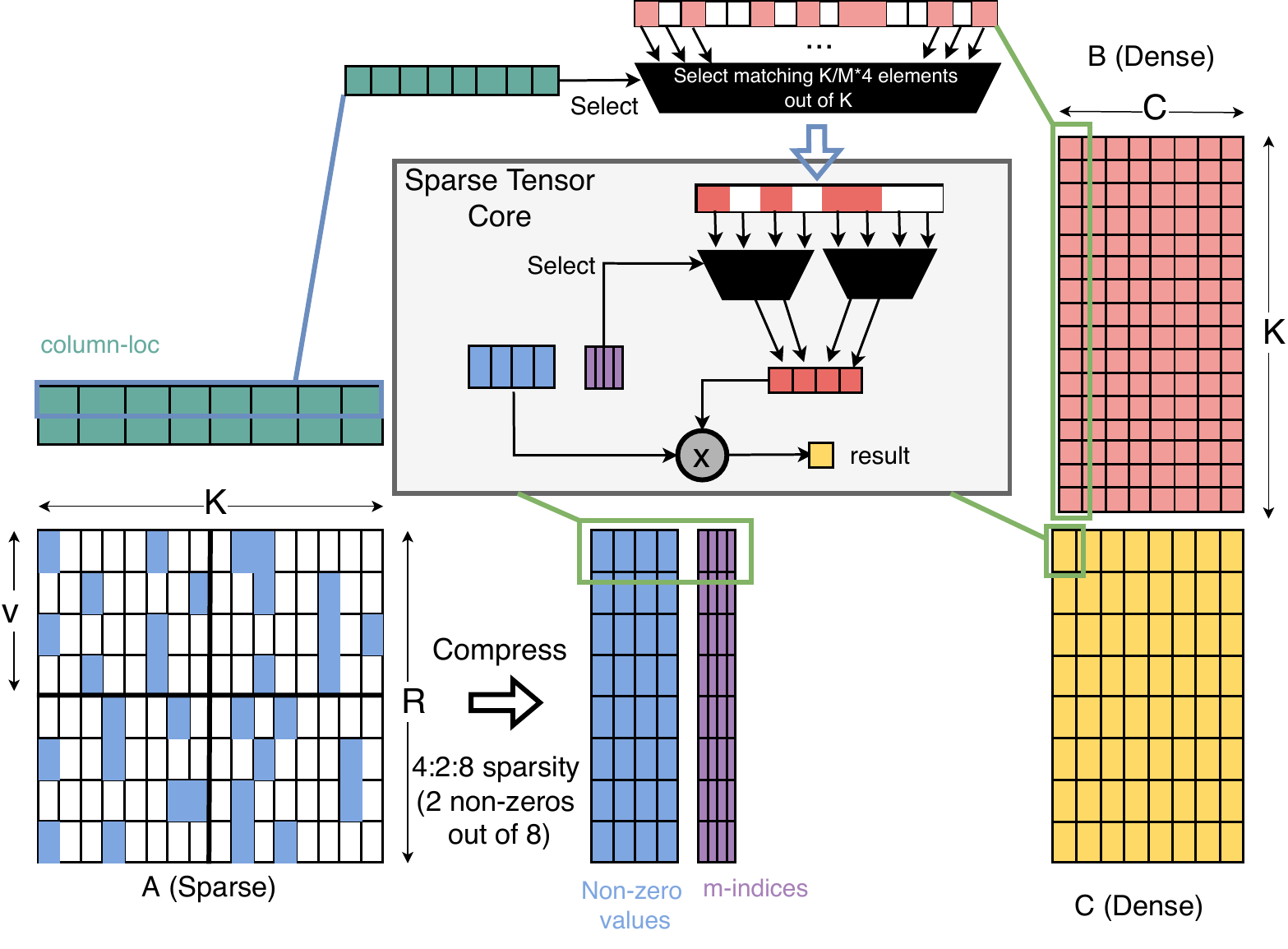}}
    \caption{Mapping a 4:2:8 format onto Sparse Tensor Core (only native support to 2:4 format)}
    \label{fig:sptcu}
\end{figure}

Figure~\ref{fig:sptcu} shows an example of how the new V:N:M format (4:2:8 in the figure) is mapped onto SPTCs, which natively only support the 2:4 format. 
It shows how the SPTC is fed with the appropriate values from a row of the sparse matrix and a column of the dense matrix.
The LHS operand is a $R \times K/4$ dense matrix after having been pruned with sparsity of $75\%$ (2:8). This pruning reduces the required multiply-and-add operations by $4$ (from 16 to 4), but also halves the rows loaded from the dense matrix B (selected by the values contained in \emph{column-loc}). 

\subsection{\textbf{Kernel design}}

\begin{figure}[ht]
    \centerline{\includegraphics[width=\linewidth]{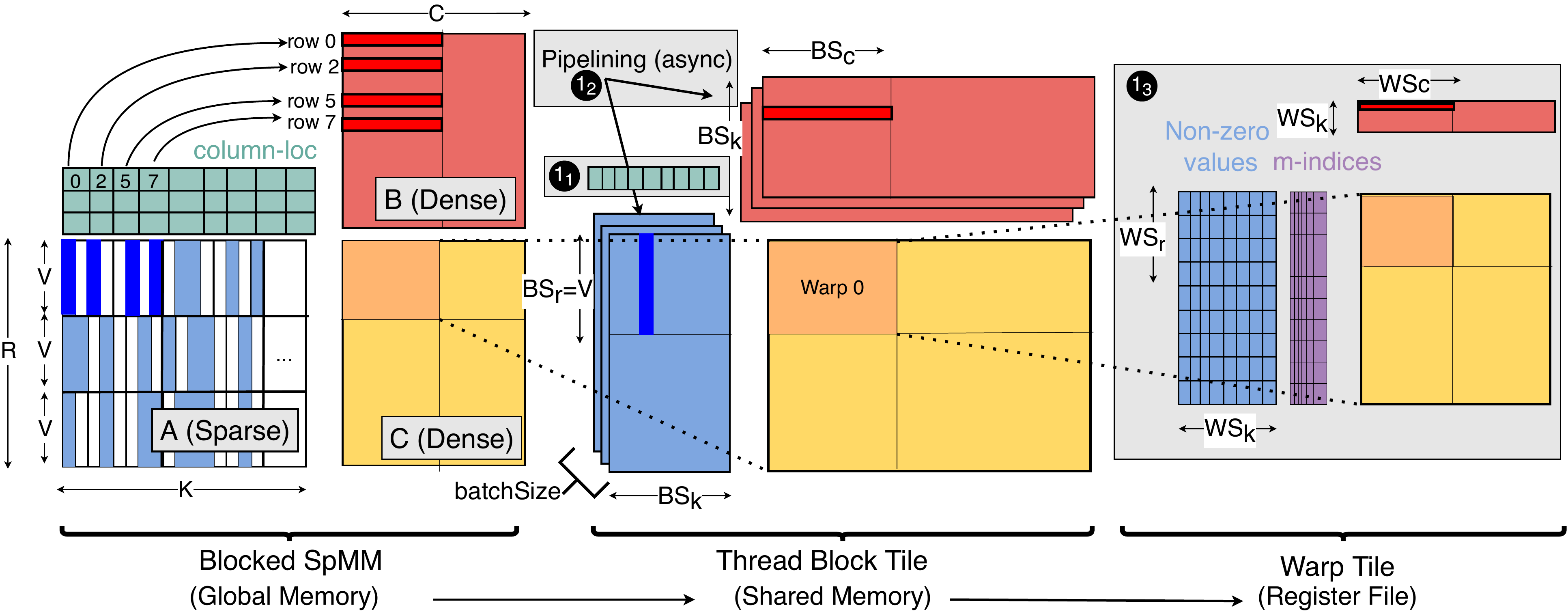}}
    \caption{Thread-Block Tile and Warp tile view (stage 1)}
    \label{fig:threadView}
\end{figure}

The design of an efficient CUDA kernel mostly depends on \textbf{three main stages}: (1) the efficient loading of the data to the top levels of the memory hierarchy (i.e., GMEM->SMEM->RF), (2) the computation, and (3) the storage of the results (i.e. RF->SMEM->GMEM).
Figure~\ref{fig:threadView} covers~\textbf{stage 1}, particularly the data movement from GMEM to RF, which is divided into 3 steps (\encircles{$1_{1}$}-\encircles{$1_{3}$}).
Figure~\ref{fig:sptcu_layout} focuses on~\textbf{stage 2}, and shows how the data in the RF is mapped onto SPTCs in three steps (\encircles{$2_{1}$}-\encircles{$2_{3}$}). Finally, Figure~\ref{fig:output} illustrates how \textbf{stage 3} is performed (steps~\encircles{$3_{1}$}-\encircles{$3_{2}$}).

Spatha is designed as a template-based library, where several parameters can be tuned depending on the input properties. Considering a $R\times K \times C$ GEMM problem, these parameters are: the thread-block tile size ($BS_{r}\times BS_{k}\times BS_{c}$), the warp tile size ($WS_{r}\times WS_{k}\times WS_{c}$), the mma instruction shape ($MMA_{r}\times MMA_{k}\times MMA_{c}$) and the level of memory pipelining (\emph{batchSize}).

\subsubsection{Stage 1-Data loading} Figure~\ref{fig:threadView} shows the Spatha procedure to load the operands from GMEM onto RF. There are two dimensions to be taken into account: the data location (i.e., GMEM, SMEM, and RF), and the scope of this data from the NVIDIA programming model perspective (i.e., thread-block, and warp).
Step~\encircles{$1_{1}$} 
loads the \emph{column-loc} structure from GMEM to SMEM with a two-level pre-fetching strategy. 
Note that the \emph{column-loc} information is used to select the rows of B to be loaded from GMEM (Figure~\ref{fig:threadView}, left side) to SMEM (step~\encircles{$1_{2}$}). 
Pre-fetching this information breaks the data dependency with the activation matrix. Furthermore, \emph{column-loc} is small, so it is convenient to load the information of multiple tiles together to maximize memory bandwidth. Next,
step~\encircles{$1_{2}$} loads the corresponding A and B tiles from GMEM to SMEM. 
Each thread-block is responsible for an output block of size $BS_{r} \times BS_{c}$. More specifically, $BS_{r}=V$, so each thread-block will load \textbf{only} the rows of B selected by the \emph{column-loc} structure. 
In order to avoid memory stalls due to data dependencies with the next steps, we pipelined step~\encircles{$1_{2}$} with step~\encircles{$1_{3}$} and stage~\encircle{2} (computation) taking advantage of CUDA asynchronous copies. The pipelining degree depends on the~\emph{batchSize} variable previously mentioned.
Finally, in~\encircles{$1_{3}$}, each warp is responsible for an output block of size $WS_{r} \times WS_{c}$, so the corresponding tiles are loaded from SMEM to RF. Emphasize that all the previously mentioned memory transactions have been optimized to use 128-bit instructions. At this point, we also load directly to the RF the \emph{m-indices} information.

\subsubsection{Stage 2-Computation}
\begin{figure}[ht]
    \centerline{\includegraphics[width=\linewidth]{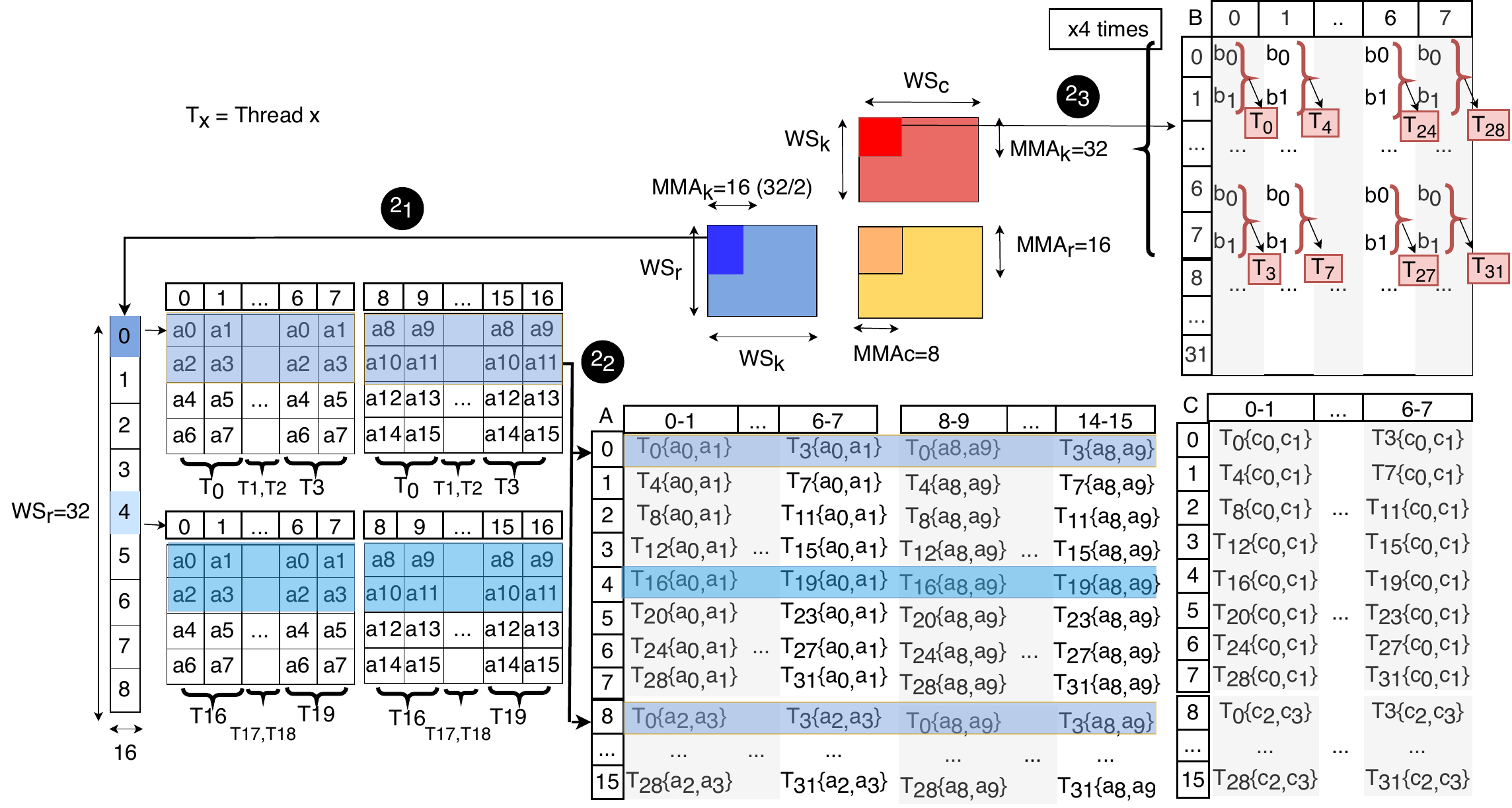}}
    \caption{SPTCs view}
    \label{fig:sptcu_layout}
\end{figure}

. When all the data is loaded in the RF, stage~\encircle{2} starts, which performs the Matrix Multiply-Accumulate (\emph{mma.sp}) on this data using SPTCs. Figure~\ref{fig:sptcu_layout} shows a detailed view of stage~\encircle{2}, depicting how the data in the RF is mapped onto SPTCs to be executed. Each warp has to break down the warp tile into instruction tiles, which depends on the instruction shapes available on SPTCs, in this example~\emph{m16n8k32}. The first step~\encircles{$2_{1}$}, selects $MMA_{k}=16$ elements from the warp tile and maps this data to SPTCs following step~\encircles{$2_{2}$} layout. This layout represents the LHS fragment to the~\emph{mma.sp} instruction. That means that, if $WS_{r}=32$, we will need to iterate twice over the rows of A's warp tile. Similarly, the next step maps the B's warp tile information into SPTCs following step~\encircles{$2_{3}$} layout, which represents the RHS fragment to the~\emph{mma.sp} instruction. At this point, the~\emph{mma.sp} instruction is executed.

\begin{figure}[h]
    \centerline{\includegraphics[width=0.9\linewidth]{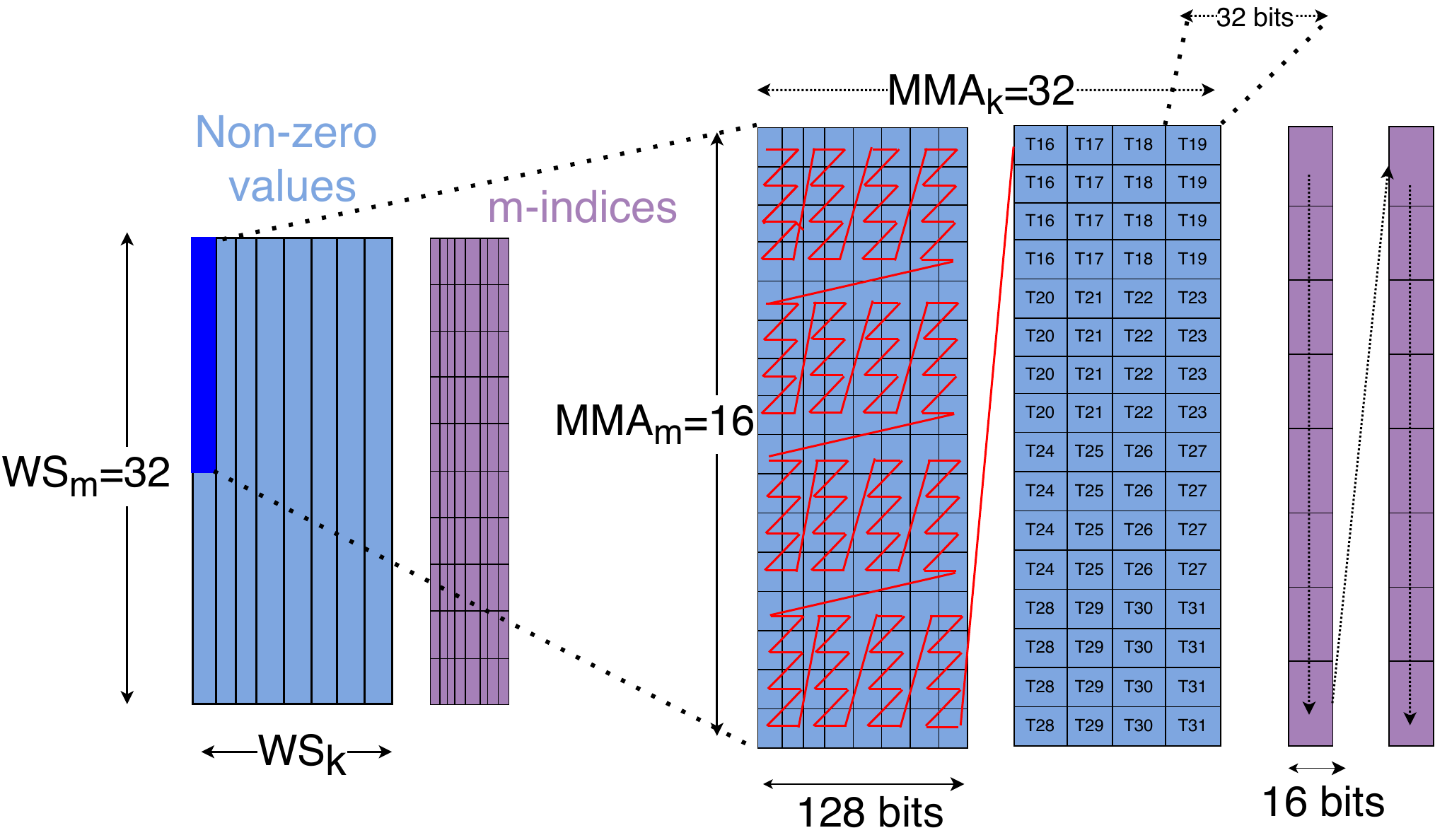}}
    \caption{Storage order}
    \label{fig:order}
\end{figure}

\paragraph{Storage order} Related to stage~\encircle{1} and~\encircle{2}, we propose a specific order to store the non-zero values and the~\emph{m-indices} structure of the V:N:M format, which merges, once again, the block-wise and the N:M principles. This order is represented in Figure~\ref{fig:order}, and it seeks to optimize the data traversal during the data loading and computation. In this representation, half of the non-zero structure shows the access pattern followed to store the data, while the other half shows how the second half-warp is mapped into this structure. This storage order enables 128-bit memory transactions, ensures memory coalescence, and can dispense with the \emph{ldmatrix} instruction, which is known to cause bank conflicts and can require more Shared Memory transactions to sequentially serve the memory access~\cite{Sun_2023}.

\subsubsection{Stage 3-Result storage} 
\begin{figure}[ht]
    \centerline{\includegraphics[width=\linewidth]{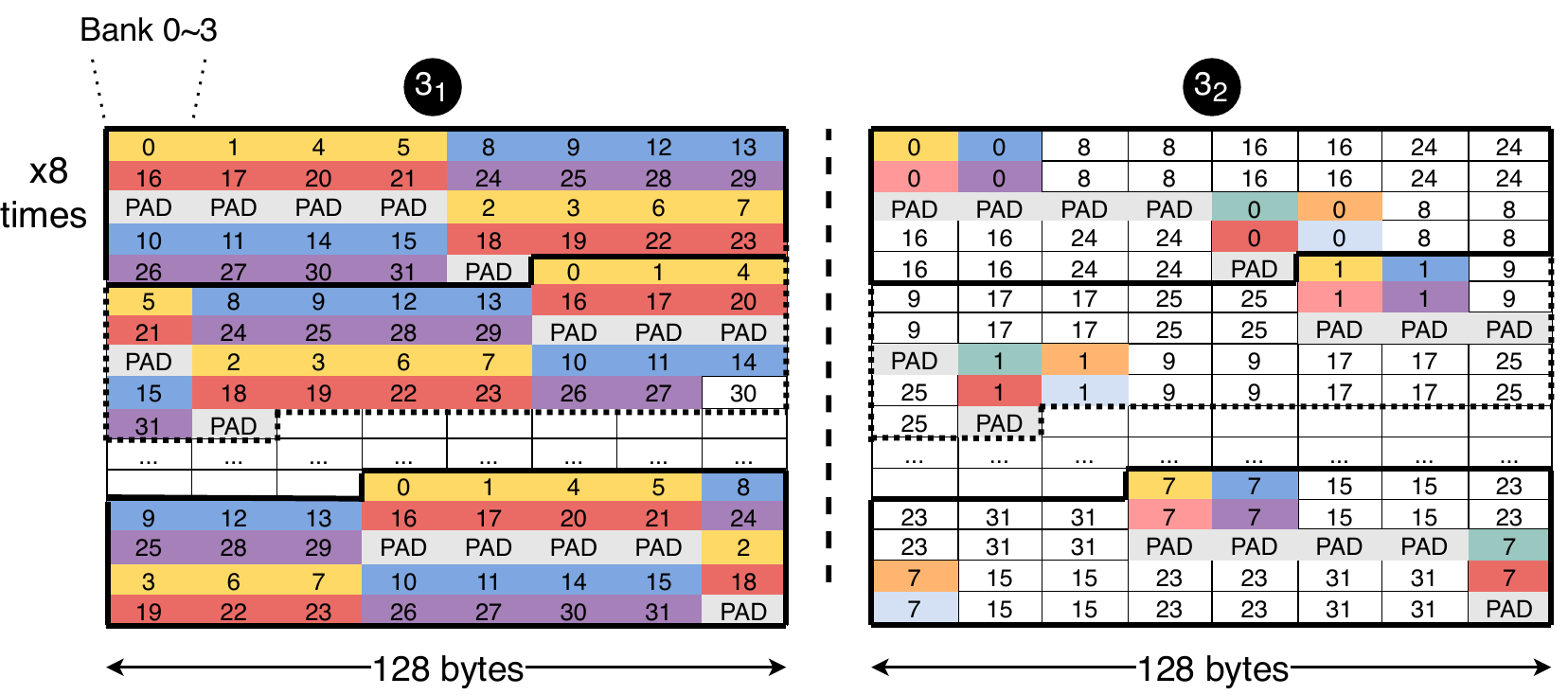}}
    \caption{Conflict-free accesses for output tiles on SMEM}
    \label{fig:output}
\end{figure}

Once the product is calculated, we have to write the output tiles back to GMEM (stage~\encircle{3}). This requires storing the intermediate partial results in SMEM. On NVIDIA GPUs, shared memory is partitioned into banks, each one of $32$ bits. Each bank can only address one position at a time, so if a quarter-warp (128-bit instructions) tries to access the same bank, the instruction will be serialized. This effect is known as bank conflict. An example of thread mapping to SMEM with $BS_{c}=64$ is shown in Figure~\ref{fig:output}. The left side of the figure shows how the threads in a warp are mapped to SMEM banks during the storage of their partial results (step~\encircles{$3_{1}$}). 
These stores are performed with 128-bit instructions. Padding elements have been added to avoid bank conflicts. In this specific example, each thread has accumulated $8$ partial results ($BS_{c}/MMA_{c}=64/8)$, so the thread mapping is repeated $8$ times, meaning that each thread needs $8$ iterations to store its partial results. Each color represents a quarter-warp, so we can see that each group of $8$ consecutive threads accesses a different memory bank in the same iteration.

The right side of Figure~\ref{fig:output} shows step~\encircles{$3_{2}$}, that is, the SMEM thread mapping designed to read the previously stored intermediate results, and finally, write them back to GMEM. The loads from SMEM and the stores to GMEM are performed with 128-bit instructions. Once again, each thread will need to access SMEM $8$ times to read all the data. We have colored the accesses related to the first quarter-warp, what depicts a conflict-free layout.

\begin{figure*}[h]
    \centerline{\includegraphics[width=\linewidth]{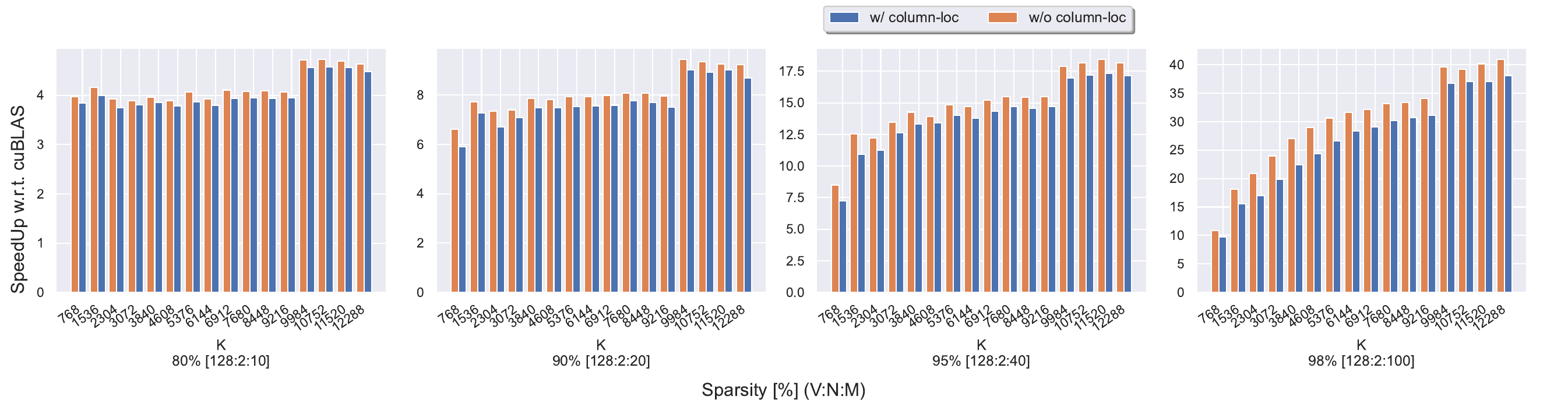}}
    \caption{Ablation study of column-loc with different sizes of the inner $K$ dimension and different V:N:M formats ($BERT_{large}$)}
    \label{fig:ideal}
\end{figure*}

\paragraph{\textbf{Ablation study - Spatha performance and column-loc overhead}} In Figure~\ref{fig:ideal}, we present the results of a microbenchmark study on matrices of fixed outer dimensions (corresponding to the size of one BERT$_{large}$ weight linear layer), but varying the inner (sparsified) one, $K$ ($1024\times K\times 4096$). The study was conducted using different sparsity levels, specified by different N:M combinations (from 2:10 to 2:100), while the vector size $V$ was kept constant at $128$. Furthermore, to measure the effect of using the \emph{column-loc} mechanism, we tested the performance with and without this structure. In the latter we used fixed indexes to simulate an ideal situation with no memory accesses.
These experiments are performed on an NVIDIA RTX 3090 GPU, equipped with SPTCs. The results show that Spatha achieves speedups for sparse computation, approaching theoretical peak performance for a given sparsity level considering the operation count reduction w.r.t. the dense counterpart version. This effect becomes more pronounced as the GEMM problem size increases, as it tend to have higher arithmetic intensity. For instance, at a sparsity level of $80\%$ (2:10 format), the speedup is approximately $4.5\times$, where $5\times$ is the ideal scenario. Then, the speedups reported are $8.5\times$, $17.5\times$, and $37\times$ for sparsity levels of $90\%$ (2:20), $95\%$ (2:40) and $98\%$ (2:100), whose theoretical caps are $10\times$, $20\times$ and $50\times$, respectively. It can be observed that, for every sparsity ratio, the~\emph{column-loc} structure's overhead has a negligible effect on the overall time, despite being a software approach to support arbitrary N:M ratios. However, the impact of~\emph{column-loc} becomes slightly more noticeable when dealing with 2:100 sparsity, which is not practical for DL applications in real-world scenarios.

\begin{figure}[t]
    \centering
        \centering
        \includegraphics[width=\linewidth]{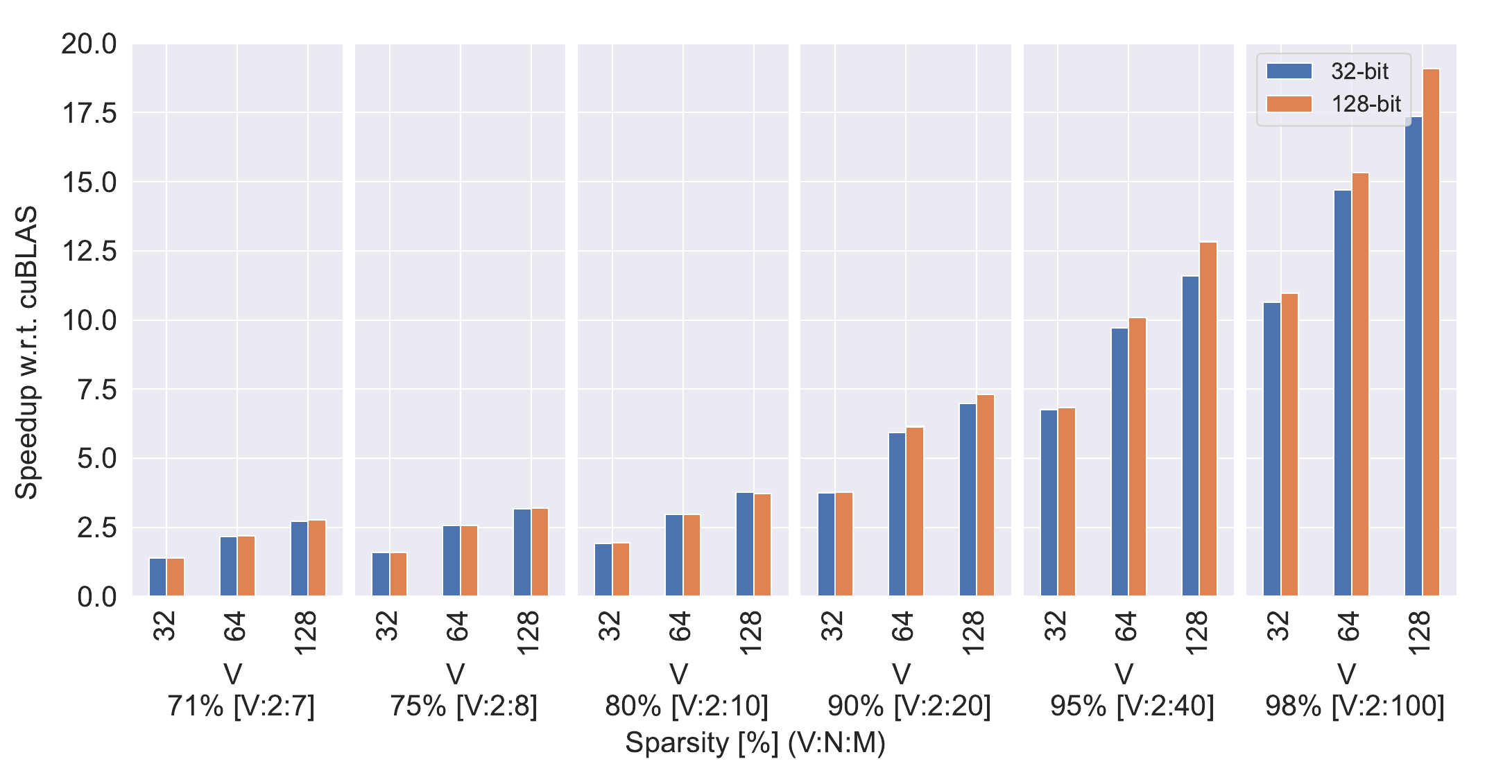}
    \caption{Scaling study of wide shared memory stores for different V:N:M configurations}
    \label{fig:abl_out}
\end{figure}

\paragraph{\textbf{Scaling study - Impact of V and output layout format}} The $V$ variable in our V:N:M format can be used to define trade-offs between performance and accuracy in the same way that the block-size in block-wise pruning, for example. 
To study this, we performed a second ablation study on one matrix from $BERT_{large}$ (size $1024\times 4096\times 4096$).
Figure~\ref{fig:abl_out} shows the performance results of Spatha on this matrix using three different vector lengths: $32$, $64$ and $128$. This test is conducted for different sparsity levels, in practice, the test explores different configurations of the V:N:M values. Furthermore, in order to study the impact of the previously proposed layout for writing back results (Figure~\ref{fig:output}), it is compared the effect of using such layout, enabling 128-bit SMEM stores instead of 32-bit ones. 
As we can see in Figure~\ref{fig:abl_out}, the difference in terms of speedups between the three selected vector lengths is noticeable, the value of $V$ being conditioned by the accuracy loss.
The effect of using 128-bit stores instead of 32-bit ones is noticeable in this problem size, bringing up to a $2\times$ difference in the final speedup. We performed a similar ablation test for a matrix of a GPT-3 model (size $36864\times 12288\times 4096$) and the effect of using 128-bit stores was attenuated, as the weight of the output phase in the total execution time is smaller.

\begin{figure}[t]
    \centerline{\includegraphics[width=\linewidth]{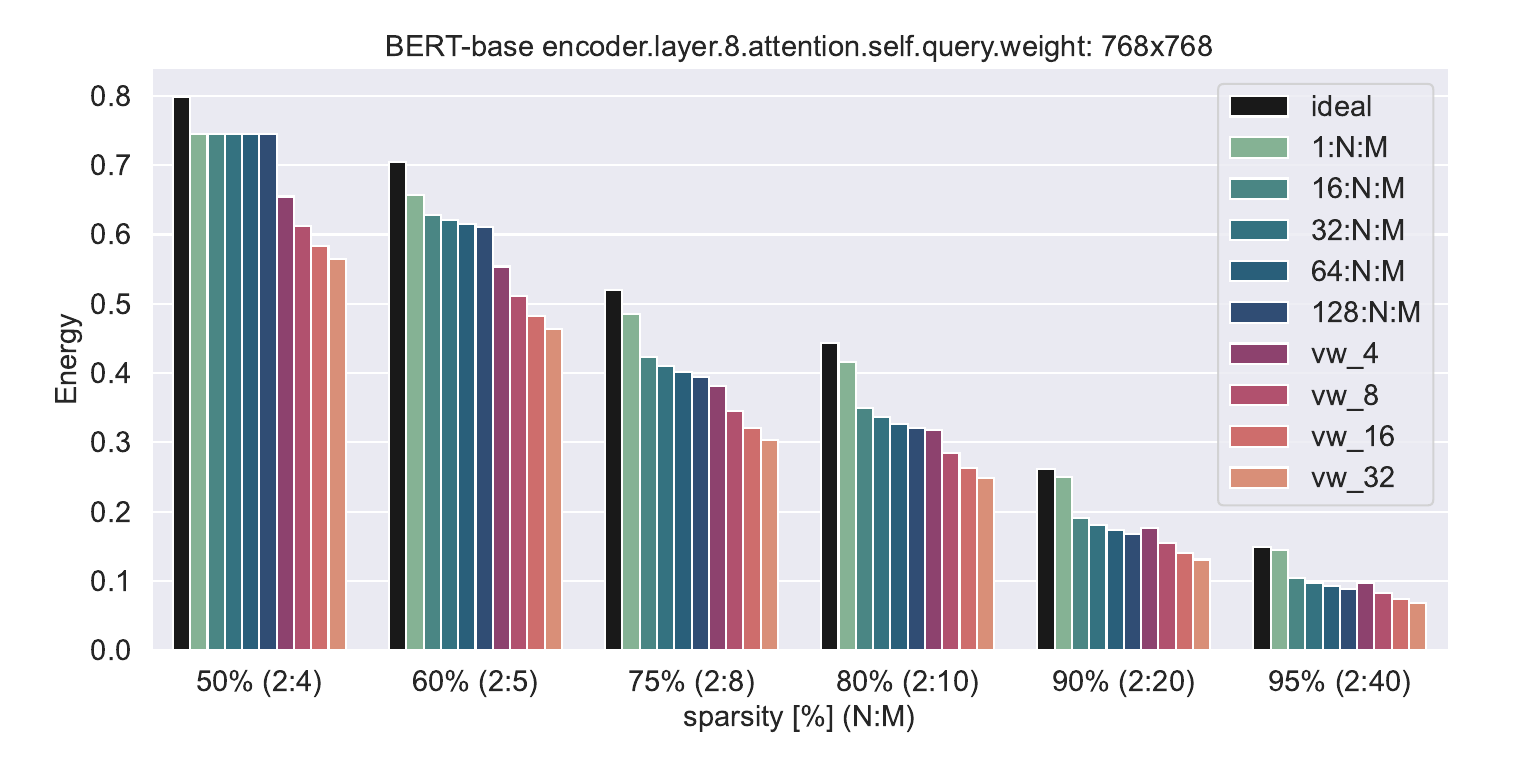}}
    \caption{Energy evaluation study on the V:N:M format}
    \label{fig:energy}
\end{figure}

\section{Energy evaluation of V:N:M}
DL pruning techniques aim to achieve the highest possible sparsity levels in the pruned models while ensuring little to no loss in accuracy. This becomes especially challenging when the target sparse format requires a specific pruning scheme, and when high sparsity levels are targeted. In these scenarios, the percentage of non-zero values is low, and their location is heavily influenced by the format. Therefore, it is crucial to demonstrate the effectiveness of new sparse formats, to ensure its applicability with minimal or no impact on accuracy.

The energy evaluation metric is employed to measure the flexibility of a format by comparing the total magnitude of the model (sum of the individual weights) before and after pruning to a specific format. Let us assume a well-optimized dense model $w^{*} \in \mathbb{R}^{d}$, where $d$ is the total number of weights. We wish to prune $w^{*}$ to a target sparsity $s\in (0,1]$ by zeroing out $s\times d$ weights. The result is a sparse model $w \in \mathbb{R}^{s\times d}$. The energy metric is defined as follows:

\[
energy = \dfrac{\sum_{i=0}^{s\times d} |w_{i}|}{\sum_{i=0}^{d} |w^{*}_{i}|}
\]

This metric yields a normalized score between $0\sim 1$, the higher the better.

Figure~\ref{fig:energy} presents the energy evaluation study for a weight tensor extracted from an encoder layer of BERT$_{base}$. This figure compares three weight selection policies: unstructured (ideal), V:N:M with different $V$ values, and vector-wise pruning with several vector lengths $l$ ($vw\_l$). The evaluation is done for different sparsity levels, whose value in the V:N:M format is controlled by the N:M ratio. 

Unstructured pruning represents the ideal non-zero selection policy, as it does not impose any restrictions on the location of non-zero values. Vector-wise pruning can accelerate sparse routines on GPUs. However, if the vector length is greater than $8$, it can significantly reduce the accuracy~\cite{10.1145/3559009.3569691,10.5555/3571885.3571934,10.1145/3458817.3476182}. The results demonstrate that the V:N:M format occupies an intermediate position between unstructured and vector-wise pruning. Moreover, it is highly robust to changes in the vector length, allowing the usage of $V=128$ while consistently preserving more energy than $vw\_8$ and $vw\_4$.

Additionally to the previous conclusions, independently of the selected pruning method, we can also see the tremendous impact on the energy of magnitude-based weight selection policies. At $50\%$ of sparsity, unstructured pruning already lost $20\%$ of the original dense matrix energy. At the other side, at $95\%$ only $20\%$ of the original energy remain in the pruned dense matrix. Thus, we can conclude that, in order to achieve moderate to high sparsity ratios in models with the dimensionality of BERT, more sophisticated pruning methods must be used. Second-order pruning offers an alternative to these problems.

\section{Second-order pruning}
Magnitude-based pruning techniques provide a straightforward approach to reducing the size of our models without requiring model evaluation for weight selection. However, while magnitude pruning can be effective at moderate sparsity levels, it becomes more challenging to select the "least significant" weights to remove when aiming for high sparsity ratios, and this can significantly impact network accuracy.

In contrast, second-order pruning methods offer a more sophisticated approach to select weight candidates for removal, by considering the difference in loss relative to the current model. Hence, they target to find the set of weights whose removal will generate a minimum loss increase. In this context, the Hessian matrix is a key component of second-order pruning methods which represents the matrix of second-order derivatives of the loss function w.r.t. the weights, mathematically expressed as $H = \nabla_{w}^{2}L$, for a twice-differentiable loss $L$. The Fisher matrix is very similar to the Hessian matrix but in the probabilistic setting, used to estimate the curvature of the loss function around the current value. As a result, this approximation allows to identify the weight parameters that have less impact in the loss function, and therefore are candidates to be pruned~\cite{10.5555/3546258.3546499}.

\subsection{The V:N:M format in 2nd order methods}
This section introduces a new second-order pruning method based on~\cite{oBERT} and tailored for the V:N:M format. This type of approach yields state-of-the-art results in LLMs for unstructured and semi-structured (block) compression. 

Let us assume we have a well-optimized dense model $w^{*} \in \mathbb{R}^{d}$, where $d$ is the total number of weights. Our target is to identify a set of weights $Q$ that we can prune with a minimum loss increase. Te following saliency score function is defined to rank groups of weights~\cite{oBERT}:

\[
\rho_{Q} = \dfrac{1}{2}(E_{Q}w^{*})^{T}(E_{Q}\widehat{F}^{-1}(w^{*})E_{Q}^{T})^{-1}E_{Q}w^{*}
\]

where, 
\begin{itemize}
    \item $\widehat{F}^{-1}(w) \in \mathbb{R}^{d \times d}$ is the Fisher matrix.
    \item $E_{Q} \in \mathbb{R}^{|Q|\times d}$ is a matrix composed of the corresponding canonical basis vectors for a set of $Q$ weights.
\end{itemize}

Thus, the set of canonical basis vectors $E_{Q}$ depends on the specific sparse format we are using. For instance, in 2:4 sparsity, the canonical vectors are:
\[ E_{Q}=[[1,\!1,\!0,\!0],[1,\!0,\!1,\!0],[1,\!0,\!0,\!1], [0,\!1,\!1,\!0],[0,\!1,\!0,\!1],[0,\!0,\!1,\!1]]
\]

As observed, $E_{Q}$ encompasses all possible correlations between $2$ weights, in a set of $4$ elements.
In general, for an N:M format, this approach requires evaluating $\binom{M}{N}$ combinations to determine the best one, which can turn into an intractable combinatorial problem. Furthermore, in the V:N:M format, the addition of a new dimension $V$ amplifies the complexity as it requires finding the optimal set of $V\times N$ weights, leading to a combinatorial explosion. 

To address these challenges, we adopt a similar approach as~\cite{oBERT} between sets of $Q$ elements, which involves disregarding correlations between rows within $V\times M$ blocks. This simplification drops the number of combinations to evaluate. Additionally, to mitigate combinatorial issues that may still arise within $1\times M$ groups, we propose a {\bf pair-wise approach} where correlations are calculated between pairs of elements, that is:
\[
E_{Q} = [[1,0], [0,1], [1,1]]
\]

Depending on the $N$ and $M$ values, we can modulate the complexity of the problem to be solved by dynamically selecting the m-combinatorial or the pair-wise approach.

\subsubsection{Gradual pruning definition}
The N:M format prunes a model to a target sparsity $s \in (0,1]$. Typically, the $s \times d$ weights are removed in one step (one-shot pruning). For $50\%$ (2:4) sparsity, this approach can be applied in most cases and the models still recover the original accuracy. However, for higher sparsity ratios, one-shot pruning reduces severely the model performance and makes hard to recover the original accuracy using additional fine-tuning steps. This negative effect on accuracy also happens in second-order methods, where one-shot pruning can result in worse Taylor approximations of the function.
We propose a structure decay scheduler for the V:N:M format, which performs N:M pruning across different $\beta$ steps, for increasing sparsity levels. This scheduler starts with a high initial value of $N_0 >> N_{\beta}$ (lower sparsity), where $N_{\beta}$ is our target $N$ value, and gradually decreases $N$ (conversely increasing sparsity) until it reaches the $N$ target value. This gradual pruning approach mitigates the adverse effects on network accuracy and improves the recovery of the accuracy in subsequent fine-tuning processes.

\section{Evaluation}
We evaluate the performance on an NVIDIA RTX 3090 GPU of the Ampere architecture equipped with SPTCs. We compare the performance of Spatha with different sparse libraries (cuSparseLt, CLASP, Sputnik) and also with a dense counterpart version (cuBLAS). We build our benchmarks on matrices from real-world LLMs. Additionally to these micro benchmarks, we also conduct a case study on real-world applications. At this point, we demonstrate the proposed second-order pruning technique, and we benchmark the end-to-end performance of Spatha on different LLM models (BERT, GPT-2, and GPT-3).

\subsection{Comparison with existing dense and sparse libraries}
\begin{figure}[ht]
    \centering
    \begin{subfigure}[t]{0.49\linewidth}
        \centering
        \includegraphics[width=\linewidth]{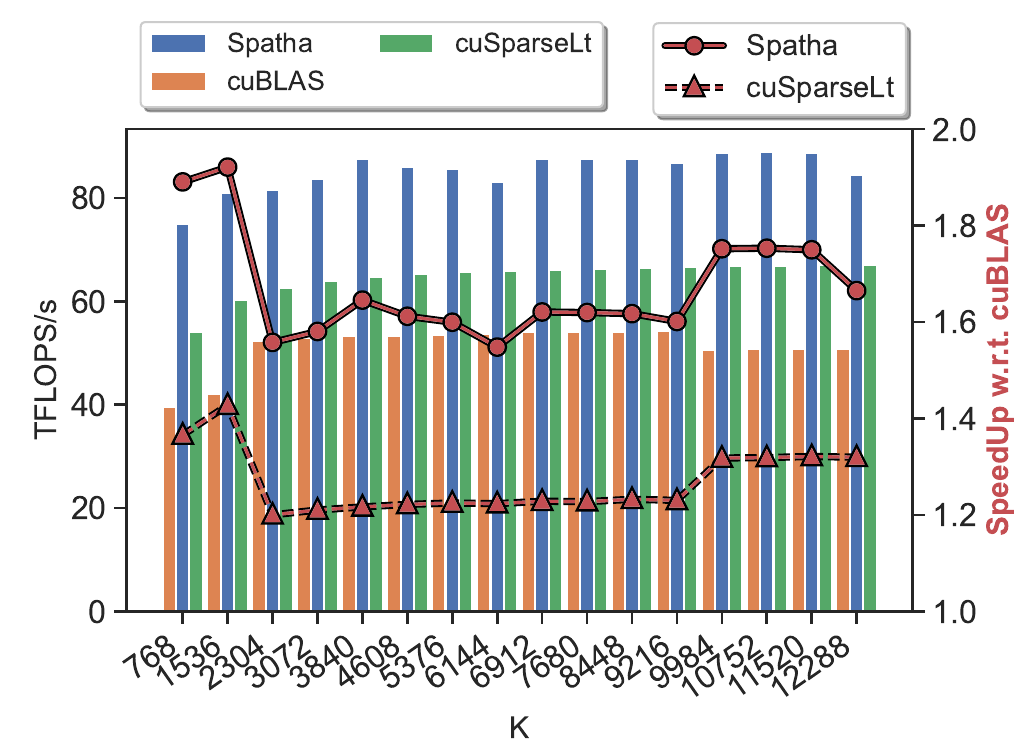}
        \caption{M=768, N=4096}
    \end{subfigure}
    \begin{subfigure}[t]{0.49\linewidth}
        \centering
        \includegraphics[width=\linewidth]{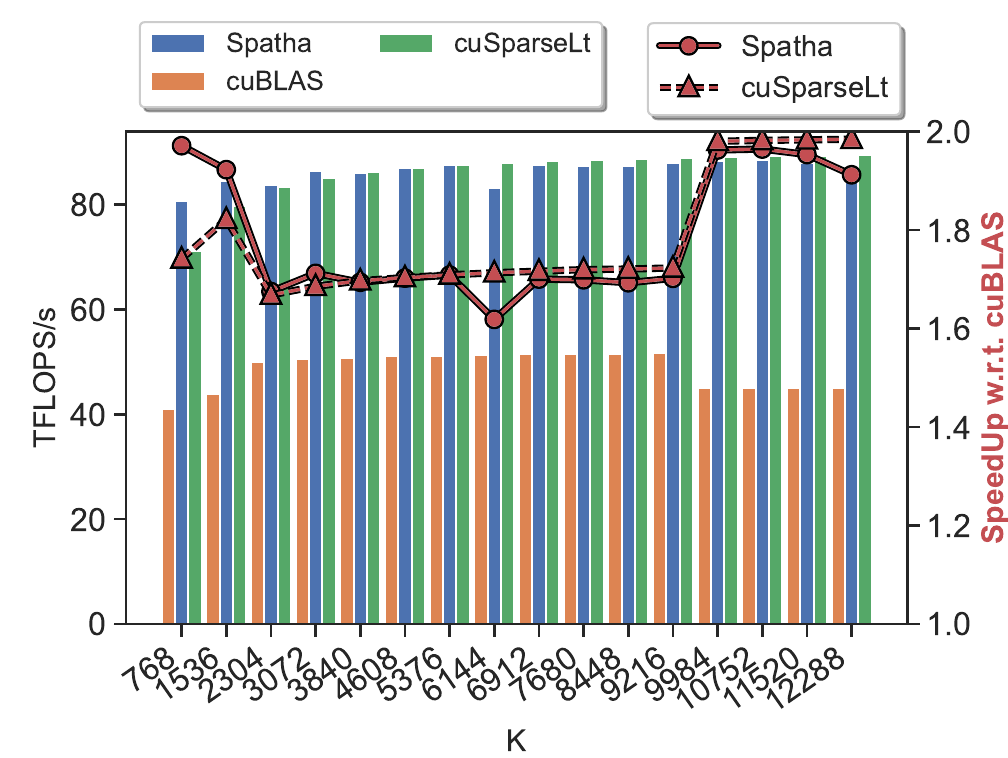}
        \caption{M=1024, N=4096}
    \end{subfigure}
    \caption{Baseline performance at 50\% sparsity (2:4 format)}
    \label{fig:cusparselt}
\end{figure}

Firstly, we evaluate our baseline implementation for 1:2:4 sparsity (50\%). Since higher N:M ratios will depend on this baseline's performance, it is crucial to have good speedup results in this configuration.
We selected cuBLAS GEMM as our dense counterpart, and for exploiting the 2:4 format on SPTCs, we used the cuSparseLt SpMM implementation, which represents the reference library on this format.
Our experiments involve varying sizes of a $R\times K \times C$ GEMM problem, where $R$ and $C$ are predetermined values from two BERT's weight linear layers ($768$ and $4096$ for BERT$_{base}$, $1024$ and $4096$ for BERT$_{large}$). 
The inner dimension $K$ of the product, which is the sparsified one, is variable in these experiments. Note that the inner dimension is usually scaled up to enhance the network accuracy. For instance, GPT-3 uses a hidden size of $12288$~\cite{brown2020language}. Figure~\ref{fig:cusparselt} reports the performance of the three contending implementations (cuBLAS, cuSparseLt and Spatha) and the speedups of the selected sparse libraries w.r.t. cuBLAS. The results show that the performance of the sparse implementation improves with the GEMM size, as larger GEMMs tend to have larger arithmetic intensity. In these microbenchmarks, BERT$_{large}$ matrices (right side) increase the computation intensity w.r.t. BERT$_{base}$ (left).
 Notably, for larger GEMM sizes, the performance of cuSparseLt and Spatha is similar, while our implementation shows better performance on smaller sizes, which constitutes an interesting feature, since Spatha can probably cover a more variety of network architectures. Overall, Spatha achieves up to $1.38\times$ speedup over the vendor library for 2:4 sparsity, cuSparseLt.

\begin{figure*}[ht]
    \centerline{\includegraphics[width=\linewidth]{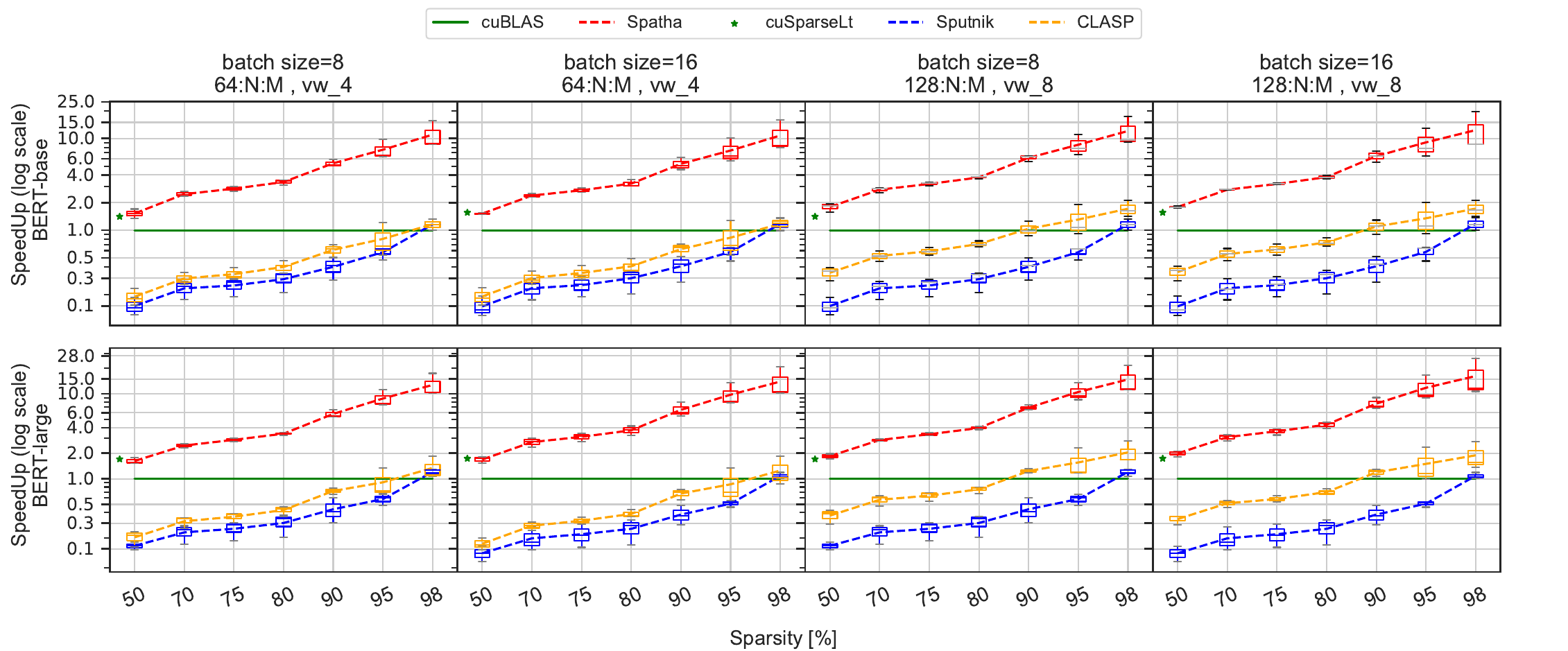}}
    \caption{Speedup results on BERT$_{base}$ and BERT$_{large}$ with sequence length=512. The notation ${V}$:N:M represents the vector length $V$ used on Spatha, while vw$_{l}$ represents the vector length ${l}$ used on CLASP. The N:M pattern related to each of the considered sparsity levels are in ascending order of sparsity: 2:4, 2:7, 2:8, 2:10, 2:20, 2:40 and 2:100}
    \label{fig:spmm_bench}
\end{figure*}

Figure~\ref{fig:spmm_bench} compares the performance of Spatha to other dense and sparse libraries for higher levels of sparsity. 
The benchmarks are built using sparse matrices from weight-pruned linear layers extracted from BERT with different sparsity levels ranging from $50\sim 98\%$. In this context, cuSparseLt SpMM implementation is the reference library to exploit the 2:4 format on SPTCs. Since there are no SpMM GPU implementations for arbitrary N:M sparsity levels, we have considered in the evaluation the following third-party libraries that support half-precision: Sputnik~\cite{gale2020sparse}, and CLASP~\cite{10.1145/3559009.3569691} which extends vectorSparse~\cite{10.1145/3458817.3476182} to the latest generations of NVIDIA GPU architectures.
While~\cite{gale2020sparse} has been designed for non-structured sparse matrices,~\cite{10.1145/3559009.3569691} is focused on semi-structured sparse input matrices following the column-vector sparse format, which supports vector lengths $l=2,4$ and $8$. This configurations has been referenced in the columns of Figure~\ref{fig:spmm_bench} with the notation $vw\_l$.

The first row of Figure~\ref{fig:spmm_bench} shows the speedup results on sparse matrices extracted from BERT$_{base}$ while the second one reports that performance on BERT$_{large}$. The y-axis is represented in a logarithmic scale to make the results more readable. First of all, existing implementations for sparse computation are usually able to outperform the dense counterpart version (e.g., cuBLAS) at sparsity levels above $80\%$. However, the speedup they can achieve is usually up to $\sim 3\times$. Furthermore, these implementations are usually designed considering as a reference sparse matrices extracted from small models (e.g., ResNet) where the left operand can be a tiny matrix (e.g., $64\times 64$)~\cite{dlmc}. That influences the SpMM design, since they can afford to load the data directly into registers, for example. But when we evaluate these implementations on medium or big matrices extracted from larger models (e.g., LLMs), the performance is even worse, and they only outperform cuBLAS at sparsity levels above $90\%$.

The fact that Spatha reaches $2\times$ speedup at $50\%$ sparsity enables the achievement of high speedups as the sparsity increases, yielding up to $27\times$ in BERT-like matrices. 
We can also appreciate that the best performance in our implementation is reached as the arithmetic intensity increases, peaking for BERT$\_{large}$ with batch size $16$.

\subsection{Case study: sparse LLMs}
LLMs have revolutioned the NLP field with their unrivaled performance in various domains. Nowadays, these models are widely used in everyday technologies, such as ChatGPT. Transformer LLMs typically consist of multiple transformer layers with self-attention.

\begin{figure}[h]
    \centerline{\includegraphics[width=0.9\linewidth]{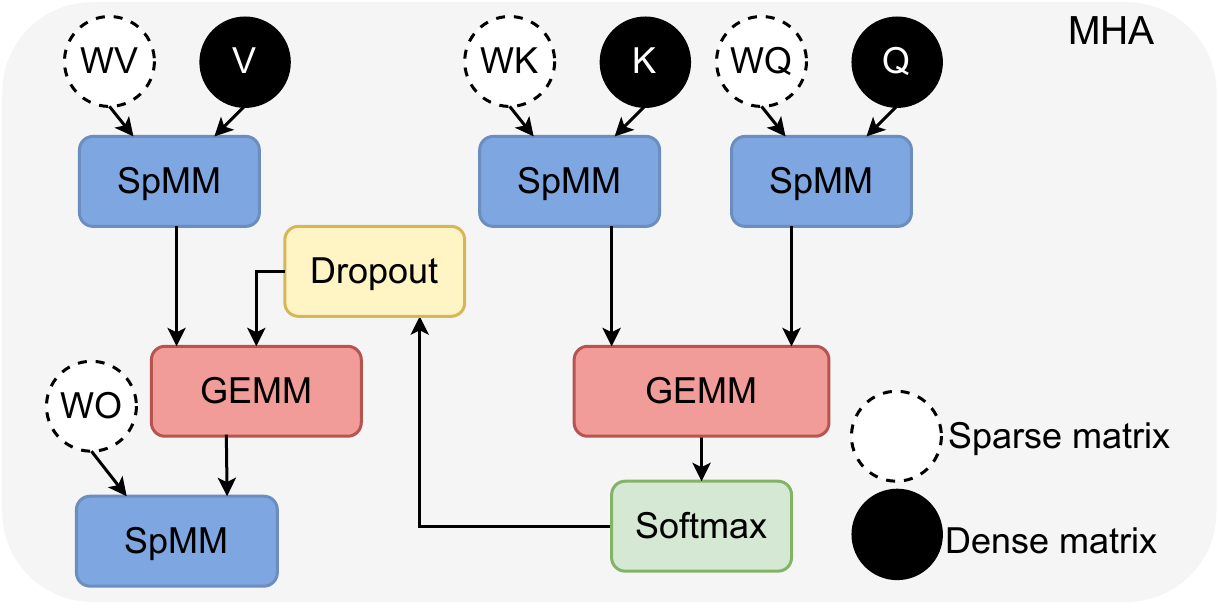}}
    \caption{Simplified view of a pruned MHA}
    \label{fig:mha}
\end{figure}

There are two major sub-components inside a transformer architecture: the multi head attention (MHA), and the fully connected feed forward network (FFN). At a higher level, the model size is determined by different configurable components, such as the head dimension, the number of heads and the number of layers, depending on the specific architecture used. 

This case study focuses on weight pruning, and explores the on computational speedups achievable with Spatha. In LLMs weight tensors are present in Linear Layers, which can be found in both the MHA and the MLP sub-components. Figure~\ref{fig:mha} illustrates a pruned MHA where four GEMM instructions are converted to SpMMs by sparsifying the corresponding weight tensors. In this study 
we demonstrate the efficiency of Spatha on different LLMs. However, it is important to note that without an efficient implementations of the SpMM instruction, the final performance of the pruned model can significantly decrease compared to the dense counterpart version.

\subsubsection{Second-order pruning at LLMs scale}
We used our 2nd order pruning approach following the V:N:M format to demonstrate its applicability to the size of LLM models. Specifically, we focused on BERT$_{base}$, one of the most commonly used LLMs, which comprises 12 transformer layers with 110M parameters. As per community standards~\cite{oBERT}, we pruned the encoder's weights of the model (85M). 
We evaluate the performance on the SQuAD v.1.1 task, which is a widely-used benchmark to measure model compression. 
Table~\ref{tab:squad} shows  the F1 score metric for different pruning techniques including: traditional N:M format (1:N:M), V:N:M format with $V$ size of $64$ and $128$, and vector-wise pruning with dense vertical vectors of size $8$ ($vw\_8$).

LLMs have been shown to be susceptible to minor model perturbations that can cause model collapse~\cite{kovaleva2021bert}. However, in these experiments we considered $75\%$ and $87.5\%$ sparsity levels, represented by 2:8 and 2:16 ratios, respectively, to demonstrate that our pruning approach produce robust results on this kind of networks.

\begin{table}[h]
\begin{center}
\begin{tabular}{ c|c|c|c|c } 
 \toprule
 Sparsity  & 1:N:M & 64:N:M & 128:N:M & vw\_8 \\ \midrule
 75\% (2:8)        &  88.61 &  88.47  &  87.94 & 88.55 \\ 
 87.5\% (2:16)     &  87.73 &  86.50  &  85.01 & 86.90\\ 
 \bottomrule
\end{tabular}
\caption{F1 score of BERT$_{base}$ on the SQuADv1.1. Dense model F1=88.43}
\label{tab:squad}
\end{center}
\end{table}

As we can see, 1:N:M, 64:N:M and $vw\_8$ slightly improve the original model accuracy at 2:8 sparsity, while the 128:N:M format presents a $0.005\%$ accuracy loss. 
For 2:16 sparsity, the four methods suffer a slight accuracy loss. Specifically, the plain $2$:$16$ format is able to recover $99\%$ of the original accuracy, while $64$:$2$:$16$ and $vw\_8$ pruning recover $98\%$. In these terms, the $128$:$2$:$16$ approach is slightly more restrictive, but is still able to recover $96\%$ of the original accuracy.

\subsubsection{Integration with Pytorch}

In order to perform the end-to-end evaluations, we have streamlined the adoption of Spatha into the PyTorch training pipeline by integrating it with the STen library \cite{sten}. This integration allows for easy addition of sparsity to existing models such as BERT and GPT with just a few lines of code. Users can specify a list of weights to be made sparse in their custom models, making the process straightforward. To facilitate this, we have defined a \texttt{VNMSparsifier} class that performs pruning while adhering to the V:N:M format constraints. Additionally, we have introduced a \texttt{VNMTensor} class that serves as a container for tensors in the V:N:M format. When using SpMM with \texttt{VNMTensor}, STen automatically dispatches it to the efficient implementation in Spatha. A pseudocode example of this integration is shown in Listing~\ref{lst:sten}.

\begin{minipage}{0.95\linewidth}
\begin{lstlisting}[caption={Pseudocode example of using Spatha and the V:N:M sparsifier},label={lst:sten},language=Python]
import sten
import spatha

@sten.register_sparsifier_implementation(
    sparsifier=spatha.VNMSparsifier,
    inp=torch.Tensor, out=spatha.VNMTensor)
def torch_tensor_to_vnm(sparsifier, tensor, grad_fmt):
    return sten.SparseTensorWrapper \
        .wrapped_from_dense(
            spatha.vnm_sparsifier(
                sparsifier.n, sparsifier.m,
                sparsifier.v, tensor),
            tensor, grad_fmt)
        
class Spmm(torch.nn.Module):
    def __init__(self, original: torch.nn.Linear):
        self.bias = original.bias
        w = original.weight.wrapped_tensor
        self.values   = w.values
        self.columns  = w.columns
        self.metadata = w.metadata
    def forward(self, input):
        return spatha.spmm(self.values, self.columns,
            self.metadata, input, self.bias, ...)    
\end{lstlisting}
\end{minipage}

\subsubsection{Sparse Inference}
We benchmark the end-to-end performance of Spatha on the inference task for different real-world LLM 
models: BERT (336M), GPT2-large (774M), and GPT-3 (175B), obtained from HuggingFace. Since GPT-3 is not a public trained model, we have created a model with the same configuration than this LLM. The target of this experiment is measuring time performance, thus, we initialized the weights of the GPT-3 model with random values. The time results on BERT and GPT2-large have been obtained over the inference of the entire model, while the results of GPT-3 were obtained by measuring the inference time of a single encoder to fit it on a single GPU.

\begin{figure}[t]
    \centering
    \begin{subfigure}[t]{\linewidth}
        \centering
        \includegraphics[width=\linewidth]{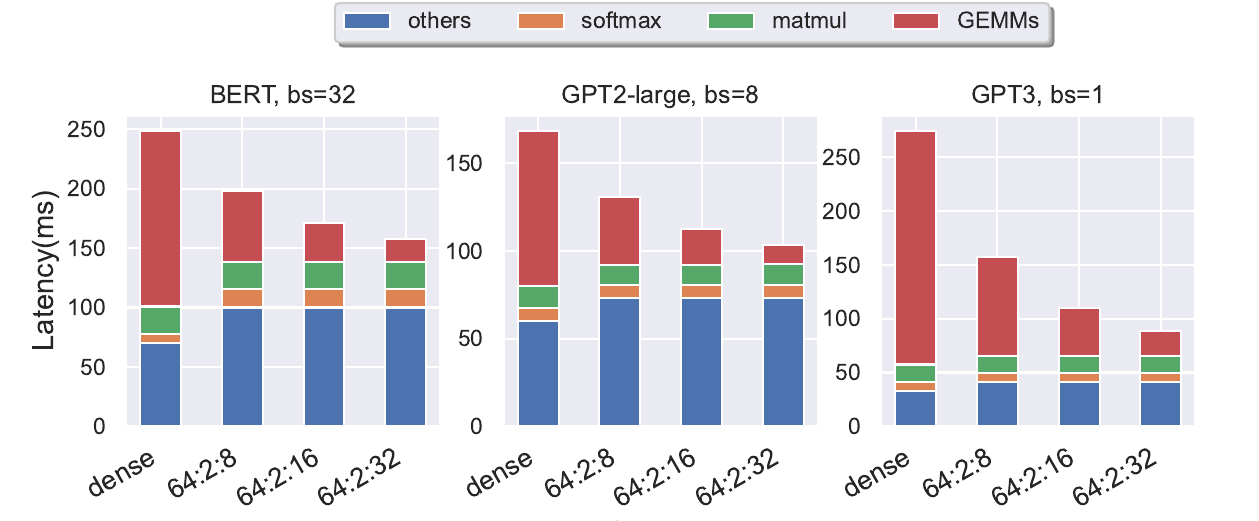}
    \end{subfigure}
    \begin{subfigure}[t]{\linewidth}
        \centering
        \includegraphics[width=\linewidth]{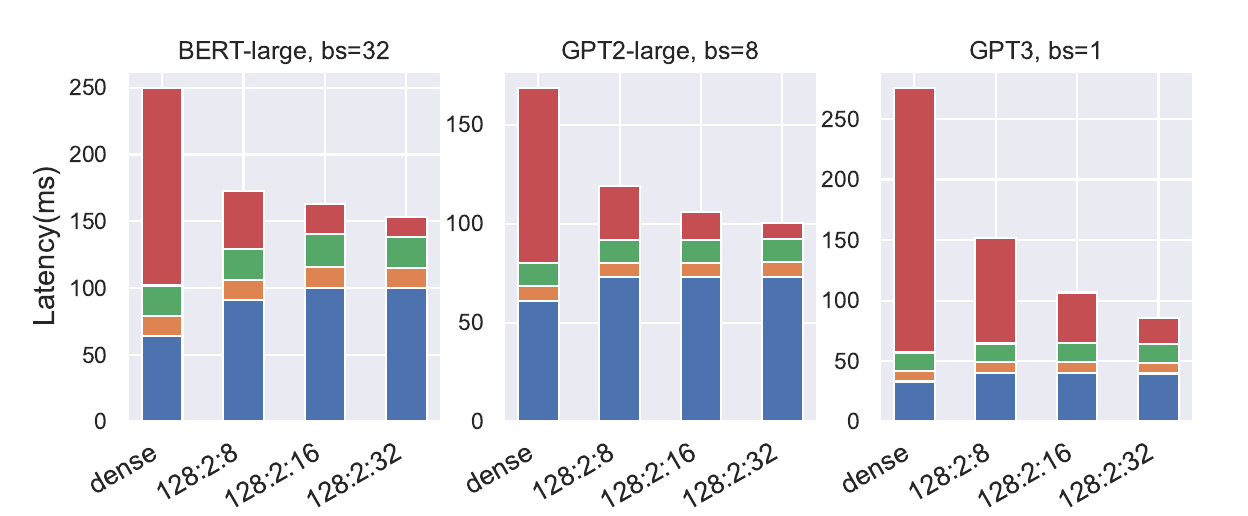}
    \end{subfigure}
    \caption{Latency of LLMs inference using Spatha}
    \label{fig:inference}
\end{figure}

Figure~\ref{fig:inference} shows the end-to-end evaluation results on the inference of these models. As we have seen in the previous micro benchmark experiments, increase the arithmetic intensity of the MMMs improves the utilization of the GPU resources, and also the final performance of the SpMM. 
We configured the three models to the larger configuration possible before achieving out-of-memory issues. In the case of BERT$_{large}$, this implied the selection of a batch size ($bs$) of 32. For GPT2-large, the $bs$ is $8$, and in the case of GPT-3, it is $1$. However, $bs$ only affect the $C$ dimension of the GEMM problem ($R\times K\times C$), while the two others, $R$ and $K$, depend on the model characteristics. Regarding these sizes, BERT has smaller weight tensor sizes (the ones sparsified) than GPT2-large, while GPT-3 is formed by weight tensors much larger than the two other models. Due to the previously described reasons, we can see that the best performance is obtained in the case of GPT-3, where the GEMM computation contributes to around $80\%$ of the total execution time.

In the case of BERT, tensor contraction time is improved up to $9.95\times$, while in terms of the whole model, the end-to-end latency is improved up to a $72\%$. For GPT2-large, the GEMM time is improved in $10.84\times$, since some weight tensors are slightly bigger, but the total GEMM time is around $50\%$, so the general improvement is limited by this factor. However, when we move to GPT-3, the tensor time contraction is improved up to $11\times$, but the GEMM time represents a much higher percentage, meaning a time reduction of up to $3.20\times$ of the total execution time of a GPT-3 encoder.

\section{Related Work}

Semi-structured pruning techniques are a hot research topic. 
The column-vector-sparse-encoding~\cite{10.1145/3458817.3476182} seeks to accelerate sparse kernels, and it achieves a speedup between $1.71\times$ and $7.19\times$ over cuSPARSE without exploiting SPTCs, and limited to the Volta architecture. 
The same authors target the SPTCs in~\cite{10.1145/3572848.3577500} proposing DFSS, a dynamic N:M sparse attention mechanism and a tailored implementation of the sparse kernels, but limited to the 2:4 format.
The unaligned group-level pruning proposed in~\cite{lee2020flexible} increases the accuracy of this kind of semi-structured pruning techniques by providing additional flexibility. 

{\emph NVIDIA cuSparse}~\cite{cusparse} is a library from NVIDIA that implements several linear algebra routines for sparse matrices stored in different compressed formats (COO, CSR and Blocked-Ellpack). It was originally created to target scientific applications.
The {\emph cuSparseLt}\cite{cusparselt} library from NVIDIA adds support for the exploitation of Sparse-Tensor Cores (SPTCs) following the N:M format, and giving support to 1:2 ad 2:4 sparsity patterns (50\% of sparsity).

{\emph Sputnik}~\cite{gale2020sparse} library has been specifically designed for DL workloads. It uses only the CSR compressed format, and it focuses on gaining flexibility on the scheduling of workloads by defining a one-dimensional tiling scheme. This library evolved to {\emph Vector-Sparse}~\cite{10.1145/3458817.3476182} adding support for the exploitation of Tensor-Core Units. It is based on using semi-structured 1D pruning, and a special compressed format called Column-Vector Sparse Encoding. As a continuation, {\emph CLASP}~\cite{10.1145/3559009.3569691} offers an SPMM implementation which extends the support of Vector-Sparse to the Ampere architecture. 

In the same line, {\emph Magicube}~\cite{10.5555/3571885.3571934} is an implementation of the SPMM and SDDMM routines for quantized sparse matrices. The kernels are complemented with en efficient online method to transpose the dense matrix.

\section{Discussion}

a) \textit{Spatha application to other tasks}. The integration of the Spatha library into STen, and the implementation of a specific 2nd order pruning technique to exploit the V:N:M format, enables distributed sparse training as a direct application of the previously mentioned contributions. Furthermore, notice that the Spatha library represents a tool to perform general Sparse Matrix-Matrix Multiplications, so can be extended to other domains other than DL.

b) \textit{Distributed deep learning systems}. In this work, we have focused on large-scale models based on LLMs. However, the Spatha library represents a generic tool for sparse MMMs. To achieve efficient large-scale DL on distributed systems, data, operator, and pipeline parallelism are often combined. In this context, Spatha can serve as a third-party implementation to accelerate the execution of these operators in the backend, and mitigate the computation bottleneck on these systems. 

\section{Conclusion}

This paper opens the possibility to use Sparse Tensor Cores (SPTCs) for arbitrary sparsity levels and N:M patterns. In order to do so, we defined a new sparse format (V:N:M), a new library to efficiently exploit the proposed kernel (Spatha), and a second-order pruning technique that demonstrated the applicability of the proposed format on real-world deep learning models.
The experiments show that this three-fold approach yields up to a $37\times$ speedup over cuBLAS at the kernel level. Furthermore, the proposed pruning technique offers a solution scalable to the dimensionality of LLMs, and is able to achieve high sparsity ratios with minimum impact in loss ($\sim 2\%$ at 2:16 sparsity on BERT models). Finally, we demonstrate the performance on end-to-end sparsity, achieving speedups on GPT-3 encoder of up to $3.20\times$ at 2:32 sparsity, what is translated into a tensor contraction improvement of up to $11\times$.

\begin{acks}
This research was supported by grants PID2019-104184RB-I00 and PID2022-136435NB-I00, funded by MCIN/AEI/ 10.13039/501100011033, PID2022 also funded by "ERDF A way of making Europe", EU; the Ministry of Education (predoctoral grant of Roberto L. Castro, FPU19/03974), by Xunta de Galicia under the Consolidation Program of Competitive Reference Groups (ED431C 2021/30), and ERC grant PSAP, no. 101002047. We also acknowledge the support from CITIC, funded by Xunta de Galicia and FEDER funds of the EU (Centro de Investigaci\'on de Galicia accreditation 2019-2022, ED431G 2019/01). Finally, we thank the Swiss National Supercomputing Center (CSCS) and the Centro de Supercomputaci\'on de Galicia (CESGA) for the use of their computers.
\end{acks}

\bibliographystyle{ACM-Reference-Format}
\bibliography{sample-base}


\end{document}